\documentclass[a4paper]{article}
\usepackage[margin=1.1in]{geometry}

\usepackage{authblk}
\usepackage{setspace}
\usepackage{array}
\usepackage{makecell}
\usepackage{amsmath, amssymb, booktabs, mathtools}
\usepackage{subfig}
\usepackage[font=small,labelfont=bf]{caption}
\usepackage{threeparttable}
\usepackage{float}
\usepackage[square, sort&compress, numbers]{natbib}

\usepackage{graphicx} 
\usepackage{hyperref} 
\usepackage[nameinlink,capitalize]{cleveref}
\hypersetup{
  colorlinks   = true,
  urlcolor     = blue,
  linkcolor    = blue,
  citecolor    = blue
}
\DeclareMathOperator{\mean}{mean}

\DeclareMathOperator{\std}{std}
\newcommand{\correspondingA}{*}

\title{Robust impact localisation on composite aerostructures using kernel design and Bayesian fusion under environmental and operational uncertainties}

\author[1]{
Dong Xiao\textsuperscript{*}
}
\author[1]{
Zahra Sharif-Khodaei
}
\author[1]{
M. H. Aliabadi
}

\affil[1]{\normalsize \slshape  Department of Aeronautics, Imperial College London, South Kensington, London SW7 2AZ, United Kingdom.}
\date{}

\begin{document}
\maketitle

\footnotetext[0]{ \textsuperscript{\correspondingA}Corresponding author}
\footnotetext[1]{ Email addresses: d.xiao21@imperial.ac.uk (D. Xiao); z.sharif-khodaei@imperial.ac.uk (Z. Sharif-Khodaei); m.h.aliabadi@imperial.ac.uk (M.H. Aliabadi)}
\footnotetext[2]{ ORCID: D. Xiao \href{https://orcid.org/0009-0006-7609-7832}{\includegraphics[scale=0.04]{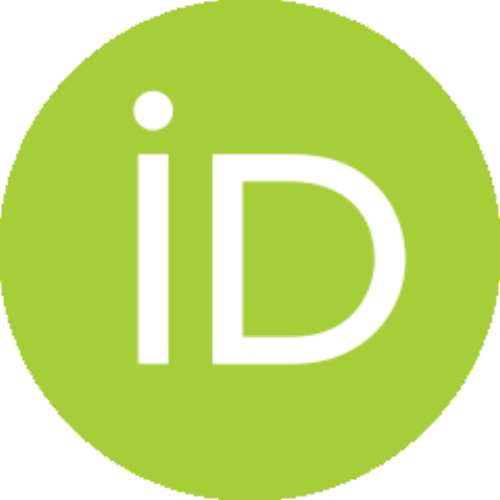}}; Z. Sharif-Khodaei \href{https://orcid.org/0000-0001-5106-2197}{\includegraphics[scale=0.04]{orcidicon.pdf}} ; M.H. Aliabadi \href{https://orcid.org/0000-0002-2883-2461}{\includegraphics[scale=0.04]{orcidicon.pdf}}
}





\renewcommand{\abstractname}{Abstract}
\begin{abstract}{\normalsize \onehalfspacing
Impact localisation on composite aircraft structures remains a significant challenge due to operational and environmental uncertainties, such as variations in temperature, impact mass, and energy levels. This study proposes a novel Gaussian Process Regression framework that leverages the order invariance of time difference of arrival (TDOA) inputs to achieve probabilistic impact localisation under such uncertainties. A composite kernel function, combining radial basis function and cosine similarity kernels, is designed based on wave propagation dynamics to enhance adaptability to diverse conditions. Additionally, a task covariance kernel is introduced to enable multi-task learning, facilitating the joint prediction of spatial coordinates while capturing interdependencies between outputs. To further improve robustness and accuracy, Bayesian model averaging is employed to dynamically fuse kernel predictions, assigning adaptive weights that account for varying conditions. Extensive experimental validation on a composite plate—including scenarios with large-mass drop tower impacts and small-mass guided drop mass impacts—demonstrates the proposed method's robustness and generalisability. Notably, the framework achieves accurate localisation without requiring compensation strategies for variations in temperature or impact mass, highlighting its suitability for real-world applications. The study also highlights the critical role of sample standardisation for preprocessing TDOA inputs, demonstrating its superiority over feature standardisation by preserving TDOA order invariance and enhancing model compatibility. These advancements establish the proposed method as a reliable and effective solution for structural health monitoring in complex and uncertain operational environments.}
\end{abstract}

\renewcommand{\abstractname}{Keywords}
\begin{abstract}{
Structural health monitoring; Impact localisation; Environmental and operational uncertainties; Time difference of arrival; Gaussian process regression; Composite kernel design and fusion}
\end{abstract}

\vspace{6pt}

\section{Introduction}
Composite aircraft airframes in operational service are vulnerable to a variety of impact events, such as tool drops, hail, and bird strikes \cite{vincent_damage_2011}. Among these, low-velocity impacts on composite aerostructures pose a significant risk of inducing Barely Visible Impact Damage (BVID) \cite{davies_impact_2004}, including matrix cracking \cite{dvorak_analysis_1987}, delamination \cite{olsson_delamination_2006}, and fiber breakage \cite{shah_impact_2019}, which can severely compromise structural integrity. To effectively detect and monitor these random impacts, passive structural health monitoring (SHM) techniques can be integrated into the airframe \cite{seydel_impact_2001}, enabling early identification of the location, presence, and severity of impact-induced damage. A comprehensive passive SHM system comprises two key components: passive sensors (hardware) permanently affixed to or embedded within the composite structures, and impact identification algorithms (software) that provide real-time localisation and estimation of impact force and damage severity. Among these processes, impact localisation is fundamental, as it provides the essential spatial information required for subsequent damage assessment and mitigation.

Impact localisation is an inverse problem aimed at determining the location of structural excitation (inputs) from structural outputs (passive sensor signals). Significant research efforts have been devoted to developing real-time impact localisation methodologies for composite aerostructures, considering their anisotropic and complex material properties. These methodologies are broadly classified into two categories: model-based methods and data-driven methods. Model-based methods rely on physical and mathematical representations of impact dynamics, including finite element analysis (FEA) \cite{yu_impact_2023, liu_quantification_2023, xiao_impact_2024, yu_accelerated_2024}, structural modal analysis \cite{goutaudier_impulse_2020, goutaudier_single-sensor_2020, zhang_efficient_2024}, and structural transfer functions \cite{el-bakari_assessing_2014, atobe_impact_2014, ciampa_impact_2012, simone_hierarchical_2019, chen_impact_2012, liu_novel_2018}. These methods excel in controlled conditions but often require precise knowledge of material properties \cite{giammaria_material_2023} and boundary conditions \cite{xu_fem-based_2023}, which can be challenging to obtain in real-world applications. Ensuring their accuracy involves iterative model updating \cite{ereiz_review_2022}, validation, and verification \cite{szabo_finite_2021}.


In contrast, data-driven methods rely on surrogate models constructed from reference databases to map structural outputs (e.g., sensor signals) to impact inputs (e.g., location and force). Traditional approaches, such as multi-layer perceptrons (MLPs) \cite{sharif-khodaei_determination_2012, ghajari_identification_2013} and Gaussian process regression (GPR) \cite{seno_uncertainty_2021, jones_bayesian_2022}, depend on features like time differences of arrival (TDOA) \cite{grasboeck_detection_2024, zeng_hybrid_2025}, rooted in wave propagation principles \cite{zhu_two-step_2018, zhu_passive_2020, meo_impact_2005, xiao_general_2025, dehghan_niri_nonlinear_2014, niri_probabilistic_2012, deng_multi-frequency_2024}. Modern deep learning techniques, including convolutional neural networks (CNNs) \cite{tabian_convolutional_2019, zhao_impact_2024}, temporal convolutional networks (TCNs) \cite{zhou_impact_2024}, recurrent neural networks (RNNs) \cite{huang_hybrid_2023, zhou_data-physics_2024}, and graph neural networks (GNNs) \cite{zhao_spatial-temporal_2023, huang_impact_2023}, directly process raw time-series data, capturing complex patterns and nonlinear relationships. However, deep learning methods often demand extensive, high-quality datasets that are challenging to obtain for real-world SHM applications.

A critical challenge in both model-based and data-driven approaches is their limited ability to generalise from controlled laboratory conditions to real-world operational environments. Real-world variability, including temperature fluctuations, impact mass, energy, and environmental noise, can significantly degrade the performance of impact localisation models. For instance, even minor temperature variations can alter wave propagation velocities, introducing errors in time difference of arrival (TDOA)-based  localisation. Furthermore, substantial changes in impact mass dramatically modify the impact responses, resulting in entirely different spectral distributions of sensor signals. While Seno \cite{seno_passive_2019, seno_impact_2019} experimentally demonstrated that small variations in TDOA occur due to changes in impact mass, height, and angle with the same impactor, these studies were limited to minor perturbations. Such findings are insufficient to address the challenges posed by significant variations in impact mass and environmental temperature, which require more robust and adaptable localisation methodologies.

To address these challenges, recent studies have explored techniques such as data augmentation \cite{hami_seno_multifidelity_2023}, transfer learning \cite{kalimullah_probabilistic_2023}, and hybrid physics-data learning \cite{xiao_hybrid_2024}. While promising, these methods face limitations in replicating the full spectrum of real-world uncertainties (data augmentation and transfer learning) or integrating physics-based and data-driven insights effectively (hybrid physics-data learning).

This study introduces a novel approach that leverages the inherent invariance in the order of TDOA under environmental and operational uncertainties. For example, temperature variations primarily scale wave propagation velocities proportionally, thereby preserving the relative order of TDOA across sensors. Additionally, substantial changes in impact mass alter the frequency spectrum of impact responses, enabling TDOA extraction at different dominant frequencies, which scale proportionally due to bending wave dispersion. Building on these insights, a composite (COMP) kernel is developed for GPR, combining the strengths of the radial basis function (RBF) and cosine similarity (COS) kernels. The RBF kernel effectively captures nonlinear relationships, while the COS kernel models linear and angular dependencies in TDOA, ensuring enhanced robustness to environmental and operational variability. To further improve adaptability, Bayesian model averaging (BMA) is employed to dynamically fuse these kernels, assigning context-sensitive weights that enhance overall model reliability and accuracy.

Extensive experimental validation was conducted on a fiber-reinforced composite panel using a drop tower machine and a guided drop mass across a wide range of impact conditions. These conditions encompassed significant environmental and operational uncertainties, including variations in impact mass, energy, angle, and temperature. The proposed method demonstrated exceptional robustness and generalisability under these diverse scenarios, highlighting its potential for real-world structural health monitoring applications. The key contributions of this study include:
\begin{itemize}
    \item Robust kernel design: Development of a novel composite kernel for GPR, integrating RBF and COS kernels. This design leverages wave propagation dynamics to enhance robustness against environmental and operational uncertainties.
    \item Bayesian kernel fusion: Implementation of Bayesian model averaging (BMA) to dynamically combine kernel predictions, enabling adaptive kernel weighting for improved accuracy and reliability across diverse impact conditions.
    \item Comprehensive experimental validation: Rigorous validation on a fiber-reinforced composite panel under a wide range of conditions, including variations in impact mass, energy, angle, and temperature, showcasing the generalisability and practical applicability of the proposed method.
    \item Sample Standardisation: Demonstration of the superiority of sample standardisation (SS) over feature standardisation (FS) for preprocessing TDOA inputs. SS preserves order invariance and enhances compatibility with GPR models under varying uncertainties.
    \item Independence from Boundary Conditions: Experimental evidence demonstrating the effectiveness of the proposed TDOA-based impact localisation method irrespective of structural boundary conditions.
\end{itemize}

The remainder of this paper is organised as follows: \cref{section passive sensing} provides an overview of passive SHM sensing for impact monitoring on aircraft structures. 
\cref{section TDOA analysis} discusses the wave propagation relationship between impact location, wave velocity, and TDOA, along with the influence of environmental and operational uncertainties, which are critical to localisation accuracy. Building on this foundation, Building on this foundation, \cref{section Gaussian process regression} introduces the GPR model, including composite kernel design and a Bayesian fusion approach. The experimental setup for impact testing under various uncertainty conditions is described in \cref{Experimental setup}. The results of TDOA extraction and impact localisation are presented and analysed in \cref{section TDOA extraction} and \cref{section: Impact Localisation}, respectively. Finally, \cref{section Conclusion and outlook} summarises the key findings and proposes directions for future research.

\section{Passive sensing for impact monitoring on aircraft structures} \label{section passive sensing}
Passive sensing offers a non-intrusive and cost-effective solution for monitoring aerostructures, providing high sensitivity to dynamic events without requiring continuous external excitation \cite{aliabadi_structural_2018}. This makes it particularly well-suited for impact monitoring in composite aerostructures, where passive sensors, such as piezoelectric transducers (PZTs), strain gauges, and accelerometers, can be embedded during manufacturing or retrofitted post-production. These sensors form a distributed network capable of continuous, real-time monitoring throughout the operational lifespan of the structure.

\begin{figure}[htb] 
	\centering
    \includegraphics[width=1\textwidth]{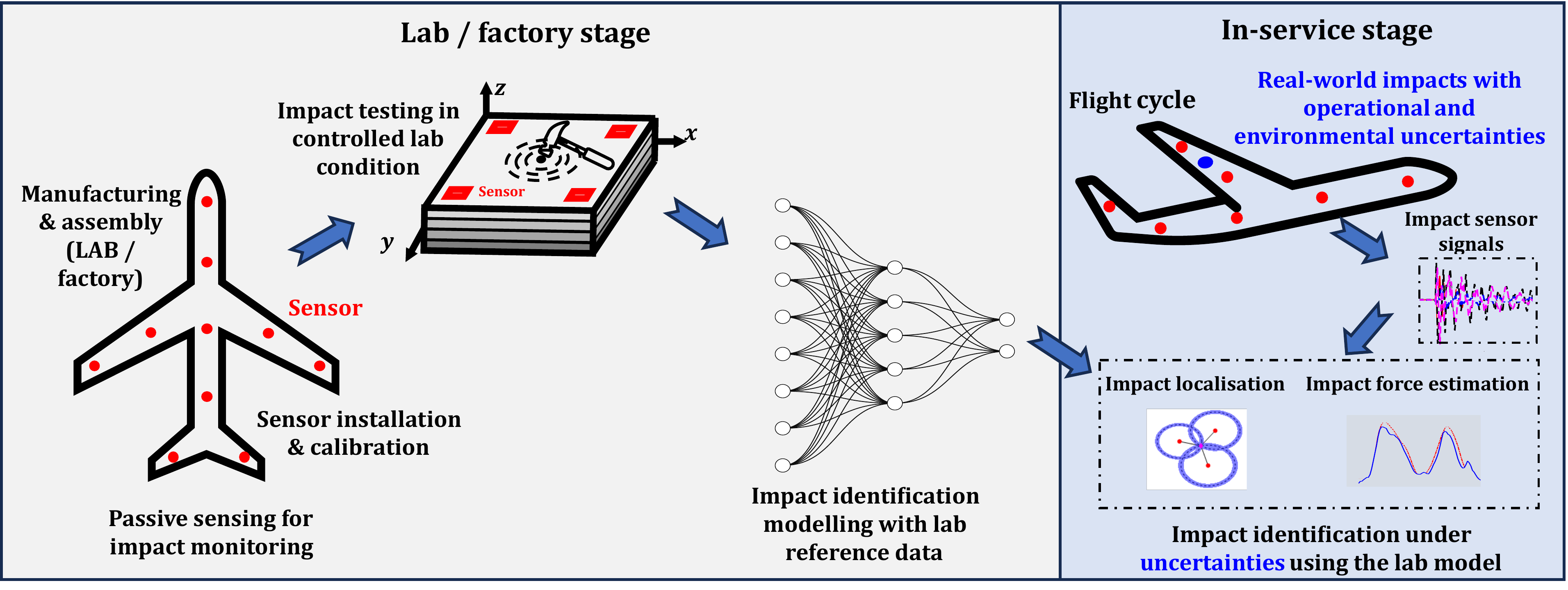} 
	\caption{Passive sensing for impact monitoring. Passive sensors capture impact responses, forming the basis for reconstructing the impact event. Accurate impact identification under environmental and operational uncertainties is essential to ensure the reliability and effectiveness of practical applications.}
	\label{FIG: Passive sensing}
\end{figure}

As illustrated in \cref{FIG: Passive sensing}, passive sensors are embedded into a composite aircraft structure during the laboratory or manufacturing stage to enable impact monitoring. When an impact event occurs on a sensorised composite structure, it generates elastic waves that propagate through the material. These waves, comprising longitudinal and flexural components \cite{moon_critical_1973, kim_impact_1979}, carry critical information about the impact event, including its location, magnitude, and duration. Passive sensors detect these waves, triggering real-time recording of sensor responses \cite{fu_event-triggered_2019}. The captured data provides the foundation for reconstructing the impact event, enabling early detection of potential damage and facilitating maintenance interventions.

Ensuring the reliability of the sensing network requires rigorous calibration and testing in both laboratory and operational environments. Laboratory calibration involves controlled impact tests at predefined locations and force levels to validate the network's sensitivity and accuracy under repeatable conditions. Field testing, on the other hand, evaluates the network’s robustness in realistic operational settings, accounting for variations in temperature, mechanical vibrations, and acoustic noise. However, extensive field testing on aircraft structures in-service is often impractical due to operational constraints, such as limited access to critical components and the difficulty of replicating controlled impacts in-service.

Given these challenges, data-driven impact localisation methods are typically trained and validated using laboratory-generated datasets, as shown in \cref{FIG: Passive sensing}. While such controlled experiments provide valuable insights and demonstrate feasibility, their applicability to real-world conditions is limited. Real-world impacts are inherently unpredictable, involving variations in parameters such as mass, velocity, angle, and environmental conditions, including significant fluctuations in environmental temperature. These environmental and operational uncertainties pose considerable challenges for data-driven models, which often rely on consistent patterns in training data for accurate predictions. This study aims to address these challenges by focusing on impact localisation under environmental and operational uncertainties, proposing robust methodologies to bridge the gap between controlled laboratory settings and real-world operational environments.

\section{Time Difference of Arrival (TDOA) analysis for impact localisation} \label{section TDOA analysis}
\subsection{TDOA relations to impact location and wave velocity} \label{section TDOA relation}
It is well established that impact events generate both in-plane waves and out-of-plane flexural waves in composite laminated structures \cite{moon_critical_1973, kim_impact_1979}. These waves are characterised by multiple modes and exhibit dispersive behavior \cite{abrate_impact_1998, moon_theoretical_1973}, where wave velocities depend on frequency. However, for low-velocity impacts, in-plane waves are typically of low amplitude and often undetectable \cite{daniel_wave_1979, tan_wave_1982}. Consequently, this study focuses on out-of-plane flexural waves, which are more prominent and carry critical information about the impact event.

The anisotropic nature of composite materials introduces a directional dependence in wave propagation, rendering the group velocity profile (GVP) both direction- and frequency-dependent. The GVP is denoted as $v_g(\theta, \omega)$ where $v_g$, $\theta$, and $\omega$ represent the group velocity, wave propagation direction and wave angular frequency. This property complicates the analysis of elastic waves but is critical for accurate impact localisation. Consider a total of $N_s$ PZT sensors distributed on the surface of the structure at coordinates $S_j=(x_j, y_j)$, where $j=1,2,...N_s$. The TDOA $\Delta t_{ij}(\omega)$ between sensor $i$ and another sensor $j$, at a given frequency $\omega$, for an impact at $P=(x, y)$ is given by \cite{kundu_acoustic_2014}:
\begin{equation}
\begin{array}{lcl} \label{EQU: TDOA relations}
\Delta t_{ij}(\omega) = t_j(\omega)-t_i(\omega) + e(\omega) = \frac{\sqrt{(x-x_j)^2+(y-y_j)^2}}{v_g(\theta_j, \omega)}-\frac{\sqrt{(x-x_i)^2+(y-y_i)^2}}{v_g(\theta_i, \omega)} + e(\omega),
\end{array}
\end{equation}
where $\Delta t_{ij}(\omega)$ represent the extracted TDOA from sensor signals, $t_j(\omega)$ and $t_i(\omega)$ are the theoretical wave arrival times at sensor $j$ and sensor $i$, respectively. $\theta_j$ and $\theta_i$ are the wave propagation direction from impact location $P$ to sensor $j$ and $i$, respectively. The term $e(\omega)$ is a random variable that accounts for uncertainties in the TDOA, arising from errors in TDOA extraction methods and inherent sensor noise. This uncertainty, $e(\omega)$, is typically modeled as a zero-mean Gaussian variable with variance $\sigma^2(\omega)$ \cite{niri_probabilistic_2012, xiao_general_2025}.

Therefore, the extracted TDOA for an impact at a given frequency $\omega$ can be represented as a vector of size $N_s$, defined as:
\begin{equation} \label{EQU: TDOA vector}
\begin{array}{lcl}
\Delta \mathbf{t}(\omega) = [\Delta t_{i1}(\omega), \hspace{1mm} \Delta t_{i2}(\omega), \hspace{1mm} \Delta t_{i3}(\omega), \hspace{1mm} \cdot \cdot \cdot \hspace{1mm} \Delta t_{iN_s}(\omega)],
\end{array} 
\end{equation}
where sensor $i$ is the anchor sensor at which the wave first arrives. The TDOA vector $\Delta \mathbf{t}(\omega)$ is inherently non-negative with at least one zero element, corresponding to the arrival of the wave at the anchor sensor.

\subsection{TDOA under environmental and operational uncertainties} \label{section TDOA uncertainties}
The reference impact dataset, collected under controlled laboratory conditions, forms the basis for training the impact localisation model. In this controlled environment, critical factors such as impact mass, velocity, impactor material, angle, temperature, mechanical vibrations, and other environmental variables are meticulously regulated and maintained at constant levels. This rigorous control ensures a consistent and repeatable dataset, enabling precise extraction of TDOA values essential for model development and validation.

Within this controlled framework, the extracted TDOA at frequency $\omega$ for an impact at location $P$, along with the corresponding group velocity profile (GVP), are denoted as $\Delta \bar{\mathbf{t}}(\omega)$ and $\bar{v}_g(\theta, \omega)$, respectively. In contrast, under real-world in-service conditions—characterized by environmental and operational uncertainties—the extracted TDOA and GVP are represented as $\Delta \mathbf{t}(\omega)$ and ${v}_g(\theta, \omega)$. These two representations highlight the discrepancies between idealised laboratory data and real-world data influenced by factors such as temperature variations, structural vibrations, and sensor noise. The ability to generalise the impact localisation model from the controlled laboratory setting to practical in-service applications fundamentally depends on the similarity between $\Delta \bar{\mathbf{t}}(\omega)$ and $\Delta \mathbf{t}(\omega)$.

While operational uncertainties, such as variations in impact mass, velocity, angle, and impactor material, do not directly alter the GVP (which depends on structural and environmental properties), they significantly influence the characteristics of the impact force and the frequency content of sensor signals. This relationship arises from the linear system property, where the frequency spectrum of the sensor signal, $\hat{s}(\omega)$, is linearly related to the impact force spectrum, $\hat{f}(\omega)$, through the system transfer function $\hat{h}(\omega)$, as $\hat{s}(\omega)=\hat{h}(\omega)\hat{f}(\omega)$ \cite{xiao_impact_2024-1}. Several studies have explored the effects of impact-related properties. Olsson \cite{olsson_mass_2000} demonstrated that the frequency content of sensor signals is primarily determined by the mass ratio between the impactor and the plate, rather than by the impact velocity. Additionally, Seno \cite{seno_impact_2019} showed that the material of the impactor significantly affects the impact force history. For instance, replacing a steel impactor head with a rubber one increases the contact time, reduces the peak force, and shifts the frequency content towards lower frequencies.

Operational uncertainties often result in substantial changes to the frequency content of impact sensor signals compared to those observed in laboratory conditions. These changes complicate the extraction of TDOA, which relies on consistent frequency components for accurate estimation. To address this, it is critical to identify a frequency $\omega$ that serves as a shared dominant frequency in both laboratory and real-world impacts. Advanced time-frequency analysis techniques, such as wavelet analysis, facilitate the identification of such shared dominant frequencies. 

In extreme cases where the dominant frequencies of laboratory and real-world impacts do not overlap, alternative strategies must be employed. One such approach involves extracting TDOA values at distinct dominant frequencies for each condition. Specifically, TDOA values can be extracted at frequency $\omega$ for laboratory impacts, denoted as $\Delta \bar{\mathbf{t}}(\omega)$, and at the dominant frequency $\omega^*$ for real-world impacts, denoted as $\Delta \mathbf{t}(\omega^*)$. 

The relationship between wave frequency and group velocity in symmetrically laminated composite plates, derived from classical laminate theory, provides crucial insights for addressing such scenarios. In the low-frequency range, the group velocity $v_g(\theta, \omega)$ exhibits a proportional relationship to the square root of wave frequency \cite{abrate_impact_1998, xiao_general_2025}, expressed as:
\begin{equation}
\begin{array}{lcl} \label{EQU: dispersion}
v_g(\theta, \omega) \propto  \sqrt{\omega}, \; v_g(\theta, \omega) = \alpha v_g(\theta, \omega^*), \; \alpha =  \sqrt{\frac{\omega}{\omega^*}}, 
\end{array}
\end{equation}
where $\alpha$ is the scaling factor. This proportionality indicates that while frequency scales the GVP, it does not alter its directional characteristics. Consequently, this scaling property extends to the extracted TDOA values. Substituting \cref{EQU: dispersion} into the TDOA expression \cref{EQU: TDOA relations}, it follows that frequency similarly scales the TDOA values without changing their relative order:
\begin{equation} \label{EQU: scaled TDOA vector}
\begin{array}{lcl}
\Delta t_{1j}(\omega^*) = t_j(\omega^*)-t_1(\omega^*) + e(\omega^*) = \alpha \left[t_j(\omega)-t_1(\omega)\right]+ e(\omega^*)  \approx  \alpha \Delta t_{1j}(\omega). 
\end{array} 
\end{equation}
where the $e(\omega^*)$ represents the uncertainty in TDOA extraction at frequency $\omega^*$. In most cases, this uncertainty is negligible compared to $\Delta t_{1j}(\omega^*)$, making the proportional scaling a reliable property for practical applications.

Environmental uncertainties, such as temperature variations, can be analysed using a similar framework. Temperature changes are assumed to induce uniform scaling in material properties and structural dimensions, affecting wave propagation speeds. This thermal effect results in a scaled GVP, with TDOA values scaled by a factor proportional to the temperature change, consistent with \cref{EQU: scaled TDOA vector}. Additionally, environmental factors such as mechanical vibrations introduce noise patterns in sensor signals. These vibrations often appear as oscillatory components within a specific frequency range, identifiable through spectral/wavelet analysis. By understanding these noise characteristics, their impact on TDOA extraction can be mitigated through filtering techniques or by selecting frequencies minimally affected by noise.

\section{Gaussian process regression with kernel design} \label{section Gaussian process regression}
\subsection{Gaussian process regression}
Gaussian Process Regression (GPR) \cite{schulz_tutorial_2018} is a Bayesian machine learning technique that excels in regression tasks due to its ability to provide both predictions and uncertainty quantification. This dual capability makes GPR particularly valuable in applications where understanding the confidence associated with predictions is as important as the predictions themselves. GPR models an unknown function $g(\mathbf{x})$ as a Gaussian process (GP), characterised by a mean function $\mu(\mathbf{x})$ and a covariance function (kernel) $k(\mathbf{x}, \mathbf{x}')$. The GP prior is expressed as:
\begin{equation}
\begin{array}{lcl} \label{EQU: GP}
g(\mathbf{x}) \sim \mathcal{GP} \left(\mu(\mathbf{x}),  k(\mathbf{x}, \mathbf{x}') \right). 
\end{array}
\end{equation} 

Given training data $(\mathit{X}, \mathit{Y})$ where $\mathit{X}$ is the input matrix and 
$\mathit{Y}$ is the corresponding output vector, and a test input $\mathbf{x}^*$, the posterior predictive distribution of $g(\mathbf{x}^*)$ is Gaussian:
\begin{equation}
\begin{array}{lcl} \label{EQU: GPR posterior}
P\left(g(\mathbf{x}^*)| \mathit{X}, \mathit{Y}, \mathbf{x}^* \right) \sim \mathcal{N} \left(\mu(\mathbf{x}^*), \sigma^2(\mathbf{x}^*) \right),
\end{array}
\end{equation} 
where the posterior mean and variance given by:
\begin{equation}
\begin{array}{lcl} \label{EQU: GPR posterior mean and variance}
\mu(\mathbf{x}^*) = k^{T}_{*}(K+ \sigma_n^2I)^{-1} \mathit{Y}, \\ [2pt]
\sigma^2(\mathbf{x}^*) = k(\mathbf{x}^*, \mathbf{x}^*) - k^{T}_{*}(K+ \sigma_n^2I)^{-1}k_{*},
\end{array}
\end{equation} 
where $K$ represent kernel matrix of training inputs $\mathit{X}$, $k_{*}$ is the kernel vector between test input $\mathbf{x}^*$ and training inputs $\mathit{X}$, $\sigma_n^2$ represents the noise variance in the observations, reflecting the model's handling of uncertain or noisy data.

The kernel $k(\mathbf{x}, \mathbf{x}')$ encodes the similarity between input points and is fundamental in defining key properties of the GP model, such as smoothness, periodicity, and response to input variations \cite{forrester_engineering_2008}. Two widely used kernels in GPR are the squared exponential kernel (also known as the radial basis function or RBF kernel) and the cosine similarity (COS) kernel. These are mathematically defined as:
\begin{equation}
\begin{array}{lcl} \label{EQU: RBF kernel}
k_{rbf}(\mathbf{x}, \mathbf{x}') = \exp{( -\frac{ \left\| \mathbf{x}- \mathbf{x}' \right\|^2} {2 l_{rbf}^2}  )}, \; 
k_{cos}(\mathbf{x}, \mathbf{x}') = l_{cos}^2 \frac{\mathbf{x}^T\mathbf{x}'}{\left| \mathbf{x} \right| \left| \mathbf{x}' \right|}. 
\end{array}
\end{equation} 
The RBF kernel is a popular choice for modeling smooth and continuous functions. It is characterised by the lengthscale parameter $l_{rbf}$, which determines the kernel's sensitivity to input differences. A smaller $l_{rbf}$ allows the model to capture finer details in the data, while a larger $l_{rbf}$ promotes smoother and more generalized predictions. The COS kernel measures the angular similarity between input vectors, ignoring their magnitudes. It computes the cosine of the angle between vectors and is scaled by the lengthscale parameter $l_{cos}$. This kernel is particularly useful when the relative orientation or proportional relationships between inputs carry more significance than their absolute values.

\subsection{Kernel design for impact localisation considering uncertainties} \label{Kernel design}
In the context of impact localisation using GPR and TDOA as input features, the kernel design is critical for addressing environmental and operational uncertainties. The kernel determines how the GPR model interprets relationships within the data, enabling it to capture patterns induced by various sources of variability \cite{satria_palar_gaussian_2020}. Specific considerations include:
\begin{enumerate}
    \item Temperature variations: Temperature changes typically cause uniform scaling in the extracted TDOA values due to the thermal expansion or contraction of the structure. To account for this, a kernel that incorporates linear components, such as the COS kernel, is essential. The COS kernel ensures the model can capture these proportional relationships, enabling it to generalise effectively across temperature-induced variations. This is critical for maintaining accuracy in dynamic operational environments.
    
    \item Frequency variability: Operational uncertainties often lead to the extraction of TDOA values at different frequencies. According to the dispersion relations in plate structures, wave group velocities scale predictably with frequency. A kernel that can model such frequency-induced linear relationships, like the COS kernel, is indispensable. This allows the GPR model to adapt to frequency-dependent TDOA variations and maintain robust performance across diverse impact scenarios.
    
    \item Simulation errors in TDOA extraction methods \cite{niri_probabilistic_2012}: Errors in TDOA extraction methods introduce non-linear variations in the extracted values. These non-linear effects, often resulting from noise or inconsistencies in the extraction algorithms, require a kernel capable of capturing such complexities. The RBF kernel is well-suited for this purpose, as it excels at modeling smooth, non-linear relationships in the data.

    \item Zero TDOA and singularity point: A singularity point is defined as a structural location where the analytical TDOA forms a zero vector—for instance, the centre of a rectangular sensor array on isotropic plates. In regions near the singularity point, TDOA values approach zero and exhibit robustness against scaling effects caused by frequency variability and temperature variations. The RBF kernel is particularly suitable for impact localisation in this area, as it relies on the magnitudes of TDOA values. Introducing an irregularly placed sensor to the array can effectively eliminate this singularity, thereby improving localisation performance.
    
    \item Multitask regression: Impact localisation involves predicting a two-dimensional output for plate-like structures, namely the spatial coordinates (x, y) of the impact location. Each coordinate pair corresponds uniquely to a TDOA vector through the GVP, as expressed in \cref{EQU: TDOA relations}. Modeling these outputs independently using separate GP may fail to capture the inherent interactions between the coordinates. Leveraging multitask regression techniques allows the model to learn and exploit similarities between the outputs simultaneously \cite{nabati_jgpr_2022}, improving accuracy and efficiency. 

\end{enumerate}
In scenarios where smooth non-linear variations and linear trends coexist—such as when environmental and operational uncertainties simultaneously affect TDOA values—a composite (COMP) kernel that integrates the strengths of the RBF and COS kernels can be effectively utilised. The composite kernel is defined as:
\begin{equation}
\begin{array}{lcl} \label{EQU: composite input kernel}
k_{comp}(\mathbf{x}, \mathbf{x}')  = k_{rbf}(\mathbf{x}, \mathbf{x}') * k_{cos}(\mathbf{x}, \mathbf{x}'). 
\end{array}
\end{equation} 
This formulation enables the GPR model to address both local, non-linear effects (captured by the RBF component) and global, proportional relationships (captured by the COS component). The RBF kernel effectively models fine-grained variations and smooth non-linearities, while the COS kernel accounts for angular similarity and linear scaling effects in the TDOA vectors. 

Since TDOA vectors are inherently non-negative, the individual kernel outputs for both COS and RBF kernels are naturally constrained within the range $[0,1]$. Without preprocessing the TDOA vectors to include negative values, the composite input kernel also remains within this range. A high composite kernel value reflects a strong similarity not only in the magnitudes of two TDOA vectors but also in their relative ordering, making it particularly suitable for capturing both amplitude and directional consistency.



To fully capture the dependencies between output coordinates, an output kernel is defined for each task ($x$- and $y$-coordinate prediction). A task covariance kernel is employed using a parameterised low-rank matrix, enabling the joint modelling of outputs. The covariance between two data points ($\mathbf{x}$, $\mathbf{x}'$) and their respective tasks ($i$, $j$) is defined as:
\begin{equation}
\begin{array}{lcl} \label{EQU: composite kernel}
k([\mathbf{x}, i], [\mathbf{x}', j])  = k_{inputs}(\mathbf{x}, \mathbf{x}') * k_{tasks}(i,j), 
\end{array}
\end{equation} 
where $k_{tasks}$ is a lookup table encoding inter-task covariances. This multitask approach effectively learns shared structures and correlations between tasks, resulting in improved localisation performance. By integrating both input and task covariance kernels, the GPR model captures a comprehensive understanding of the relationships in the data, delivering robust and accurate predictions under diverse environmental and operational conditions. 

In this study, GPyTorch \cite{gardner_gpytorch_2018}, a Gaussian process library built on PyTorch \cite{paszke_pytorch_2019}, is utilised to construct GPR models incorporating the defined input and task kernels. The hyperparameters of the GPR models are optimised using the Adam algorithm, with a learning rate of 0.1. The maximum number of optimisation iterations is set to 5000, which has been verified as sufficient to ensure convergence during training.



\subsection{Bayesian Model Averaging (BMA) for kernel fusion}
Given the availability of the aforementioned input kernels, an important question arises: which kernel is most effective for impact localisation? The answer largely depends on the conditions surrounding the target impact $x^*$ to be located. As training data are typically collected under controlled laboratory conditions, it is crucial to evaluate each kernel's performance under varying uncertainties. If the target impact is influenced by environmental variations, such as changes in impact mass or temperature, the COS kernel is expected to deliver higher localisation accuracy. Conversely, if the target impact is dominated by random noise, the RBF kernel tends to perform more reliably.

To adaptively determine the suitability of each kernel for the target impact $x^*$, two key indicators can be utilised. First, the goodness-of-fit of the GPR model on the training data can be evaluated using the marginal likelihood (ML). Second, the predictive confidence for $x^*$ can be assessed through the predictive variance $\sigma^2(x^*)$. These indicators provide complementary perspectives for kernel evaluation, facilitating the selection of the most appropriate kernel under specific conditions.

Rather than selecting a single kernel and discarding the rest, Bayesian Model Averaging (BMA) \cite{wasserman_bayesian_2000} offers a robust framework for combining predictions from multiple kernels. Each kernel is weighted according to its posterior probability:
\begin{equation}
\begin{array}{lcl} \label{EQU: BMA}
P(k_i| \mathcal{D}) = \frac{P(\mathcal{D}|k_i) P(k_i)}{\sum_j P(\mathcal{D}|k_j) P(k_j)}, 
\end{array}
\end{equation}  
where $k_i$ represents the $i$-th kernel, $P(k_i| \mathcal{D})$ is its posterior probability, and $P(\mathcal{D}|k_i)$ is its marginal likelihood. Assuming equal priors $P(k_i)$ in the absence of prior knowledge, the marginal likelihood-based weight is defined as:
\begin{equation}
\begin{array}{lcl} \label{EQU: marginal likelihood weight}
w^{ml}_i =  \frac{P(\mathcal{D}|k_i)}{\sum_j P(\mathcal{D}|k_j)}, 
\end{array}
\end{equation}  
representing the relative importance of each kernel based on its performance on the training data. To incorporate predictive confidence, a second weight can be defined as inversely proportional to the predictive uncertainty of $x^*$:
\begin{equation}
\begin{array}{lcl} \label{EQU: uncertainty weight}
w^{unc}_i =  \frac{1/\sigma_i^2 (x^*)}{\sum_j 1/\sigma_j^2 (x^*)}. 
\end{array}
\end{equation}  
By combining the marginal likelihood weight and the confidence weight through a weighted product, the final kernel weights are obtained as:
\begin{equation}
\begin{array}{lcl} \label{EQU: combined weight}
w_i =  \frac{w^{ml}_i \cdot w^{unc}_i}{\sum_j w^{ml}_j \cdot w^{unc}_j}. 
\end{array}
\end{equation}  
This formulation balances the model's fit to the training data with its confidence in predicting the target impact.

The final prediction for the target impact $x^*$ using BMA is then calculated as a weighted sum of the predictions $\mu_i(x^*), \sigma^2_i(x^*)$ from all kernels: 
\begin{equation}
\begin{array}{lcl} \label{EQU: weighted prediction}
\mu_{BMA}(x^*) = \sum_i w_i \cdot \mu_i(x^*), \; \sigma^2_{BMA}(x^*) = \sum_i  w_i \cdot \sigma^2_i(x^*).
\end{array}
\end{equation}  
This approach leverages the strengths of multiple kernels to provide an adaptive and robust localisation estimate under various uncertainties.

\section{Experimental validation} \label{Experimental setup}
Experimental impact tests were performed to collect reference impact data under laboratory conditions. These impact testing were conducted on a sensorized fiber reinforced composite plate, using three different methods of impacting the structure: drop tower, small hammer, and guided drop mass, to introduce variability in the data. The composite plate is manufactured from M21/T800 material and has dimensions of 290 × 200 × 4 mm with a layup of $[0/+45/-45/90]_{2s}$. To capture the impact-induced structural responses, a total of 6 PZT sensors (PIC 255) were strategically bonded to the surface of the plate. Detailed material properties of the plate and PZT sensor can be found in the reference \cite{seno_passive_2019, xiao_hybrid_2024, xiao_general_2025}. Based on the material properties and layup, the critical load representative of delamination onset can be calculated following \cite{suemasu_multiple_1996, olsson_mass_2000}:
\begin{equation}
\begin{array}{lcl}
F_{cr} = \pi\sqrt{\frac{32D^*G_{IIc}}{3}},\hspace{1mm} D^*\approx \sqrt{D_{11}D_{22}(A+1)/2},\hspace{1mm} A=(D_{12}+2D_{66})/\sqrt{D_{11}D_{22}}, 
\end{array}\label{EQU: delamination growth}
\end{equation}
where $G_{IIc}$ denotes the mode II interlaminar toughness, and $D^*$ represents the effective plate stiffness. Numerical and experimental findings indicate that the critical delamination force in \cref{EQU: delamination growth} is independent of the boundary conditions \cite{olsson_analytical_2001}, making it applicable to various structures. Consistent with prior research \cite{thiene_effects_2014}, this study assumes $G_{IIc}=700 \text{ J/m}^2$. The critical delamination load for the experimental composite plate is determined to be 4524.3 N. Notably, impacts with a peak force in the order of several thousand Newtons can induce delamination in the composite plate.

\subsection{Impact testing using drop tower and small hammer}
\begin{figure}[htb] 
	\centering
		\includegraphics[width=0.75\columnwidth]{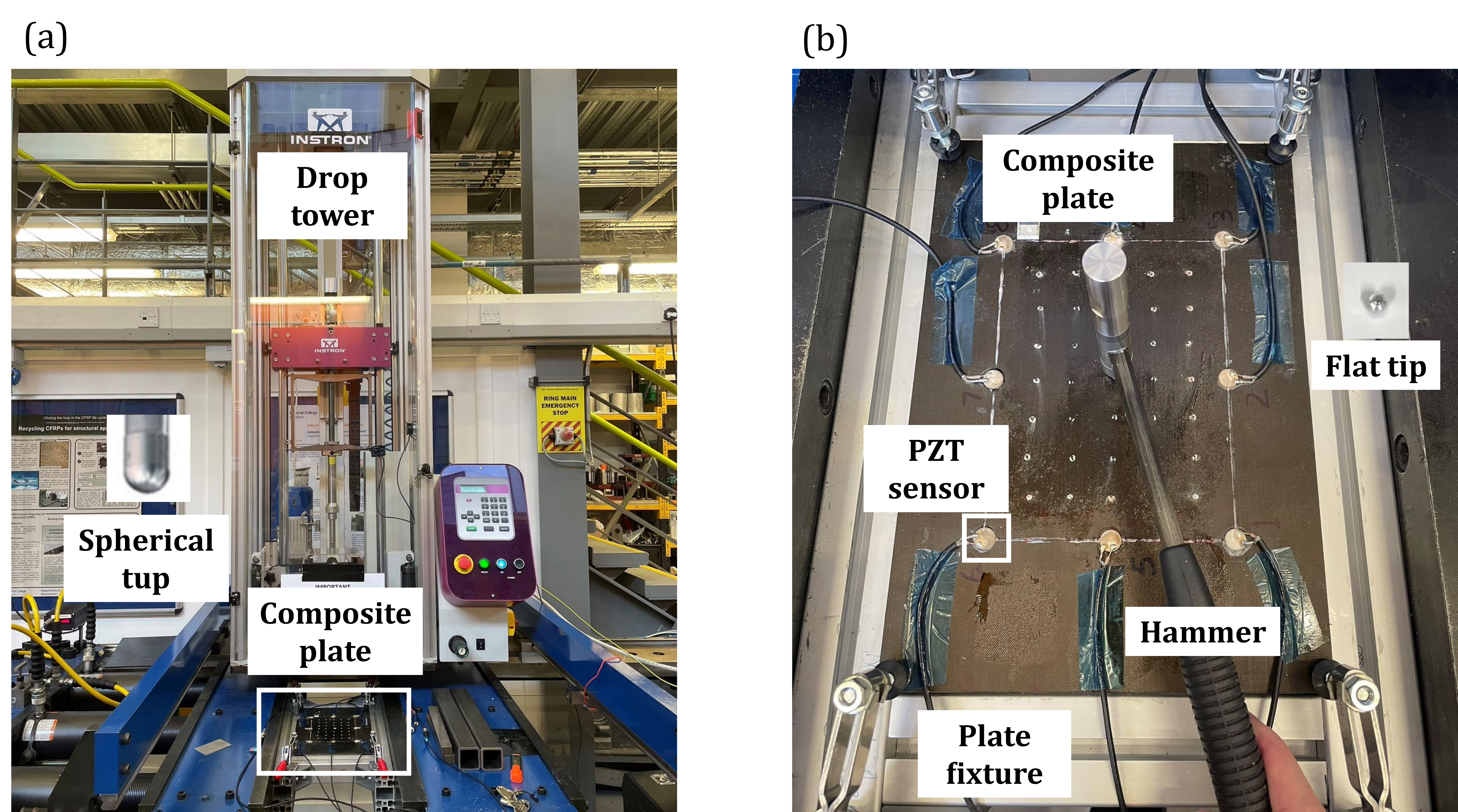}
	\caption{Experimental setup: (a) drop tower impact testing, (b) small hammer impact testing. The primary uncertainties investigated were impact mass and impactor type, given their significant influence on the spectral characteristics of the sensor signals.}
	\label{FIG: impact testing}
\end{figure}

To simulate large-mass impacts with peak forces approaching the delamination threshold, a drop tower machine (Instron-CEAST 9350) was used. The impactor was a spherical mass of 5.5 kg and 2 cm diameter. The impact energy was precisely controlled at 6 J by adjusting the drop height. For comparison, small-mass impacts were generated using a handheld hammer (PCB Piezotronics 086C03) at the same location. The hammer, which weighed 160 g, had a flat tip with a diameter of 0.63 cm. The experimental setup, including the composite plate, drop tower and handheld hammer, is illustrated in \cref{FIG: impact testing}.

\begin{figure}[htb] 
	\centering
		\includegraphics[width=0.45\columnwidth]{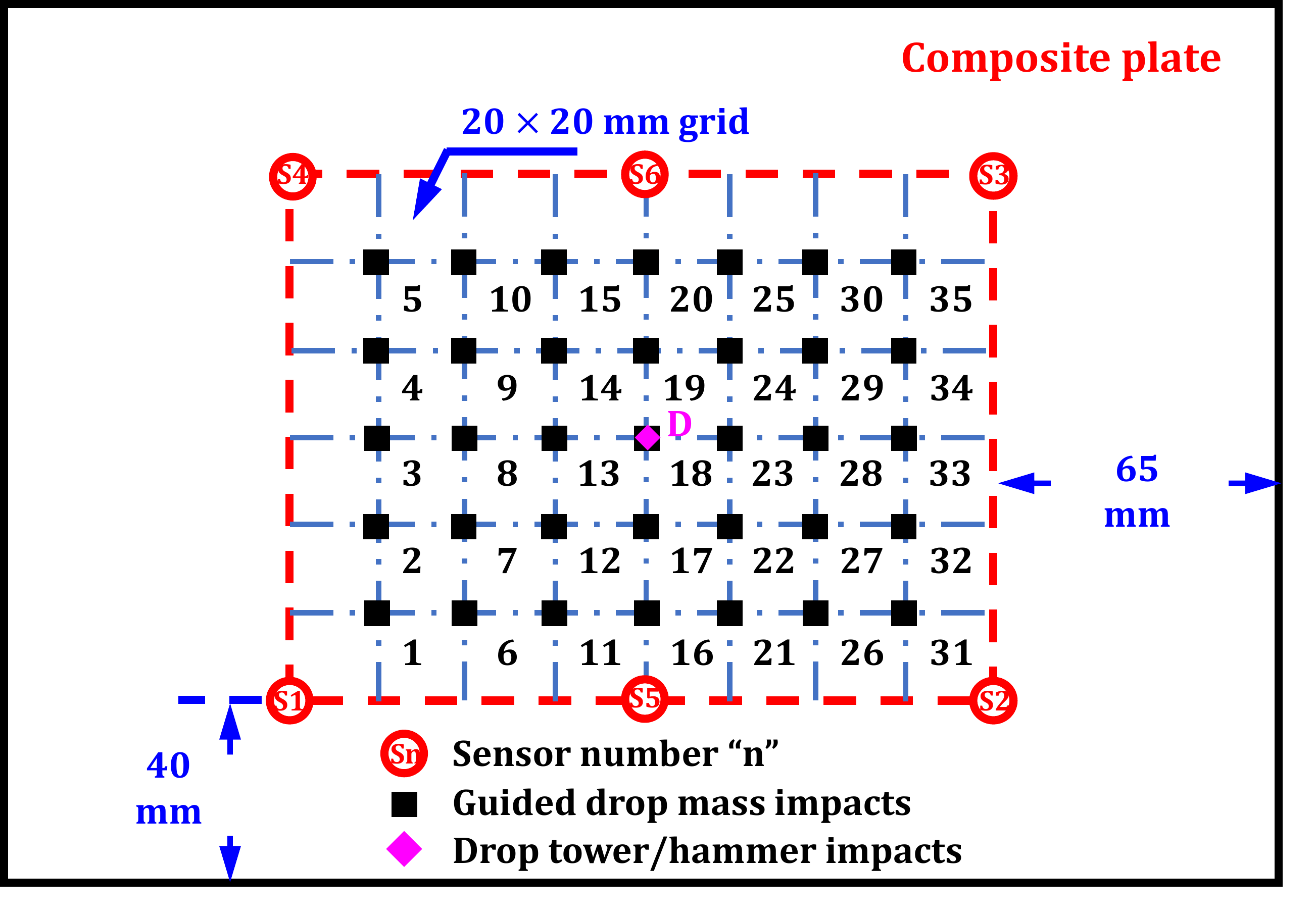}
	\caption{Layout of the composite plate used for impact testing. Six PZT sensors, numbered S1 to S6, are installed on the surface of the plate, forming a rectangular arrangement. Drop mass or hammer impacts are applied at the centre of the rectangle, while 35 uniformly distributed locations, marked as solid squares, indicate the positions for guided drop mass impacts. These will be further detailed in the following subsection.}
	\label{FIG: plate_illustration}
\end{figure}

The composite plate was supported along its two longitudinal edges and clamped at four rectangular corners using fixtures beneath the drop tower, as depicted in \cref{FIG: impact testing}. Impact-induced wave signals were captured during the experiments using four PZT sensors positioned at the vertices of the rectangle, labeled S1 to S4 in \cref{FIG: plate_illustration}. These signals were recorded using an 8-channel PXI-5105 oscilloscope with a sampling frequency of 200 kHz. The sensors were connected to the oscilloscope through 100x attenuation probes to ensure high-fidelity data acquisition. For clarity, \cref{FIG: plate_illustration} presents the layout of the composite plate, including the sensor placements and the impact location of the drop tower and hammer, marked as 'D'.
Impacts at location D were conducted successively using the drop tower at 6 J and the handheld hammer. Following each impact, ultrasonic scans were performed to assess potential impact-induced damage. The results confirmed that the composite plate remained undamaged in all tests. A summary of the conducted impact cases is presented in \cref{TB: exp impact cases}, where large-mass drop tower impacts and small-mass hammer impacts are labeled as 'DT' and 'HA,' respectively. The contact duration for these tests was derived from the impact force histories. The results indicate that large-mass impacts, with a peak force of 2143.8 N, exhibit longer contact durations and predominantly excite low-frequency response modes. In contrast, small-mass hammer impacts generate shorter contact durations, with a peak force below 200 N, and predominantly excite high-frequency modes. Notably, for hammer impacts, the contact duration in \cref{TB: exp impact cases} corresponds to the first peak in the impact force profile, which often exhibits multiple sine-like peaks. This experimental design facilitated the assessment of how impact mass and impactor influence signal characteristics, providing valuable insights into the structural response under varying conditions.
\begin{table}[ht]
\centering
\caption{Impact testing conducted using drop tower and small hammer, accounting uncertainties in impact mass and impactor material. }\label{TB: exp impact cases}
\begin{tabular}{ccccc}
\toprule
Impact case  & Impact mass (g) & Peak force (N) & Contact duration (ms) & No. of impacts \\
\midrule
DT  & 5500 & 2143.8  & 13.66 & 1 \\
HA  & 160  & 195.2  & 0.56  & 1 \\
\bottomrule
\end{tabular}\\
\end{table}

\subsection{Impact testing using guided drop mass}
\begin{figure}[htb] 
	\centering
		\includegraphics[width=0.8\columnwidth]{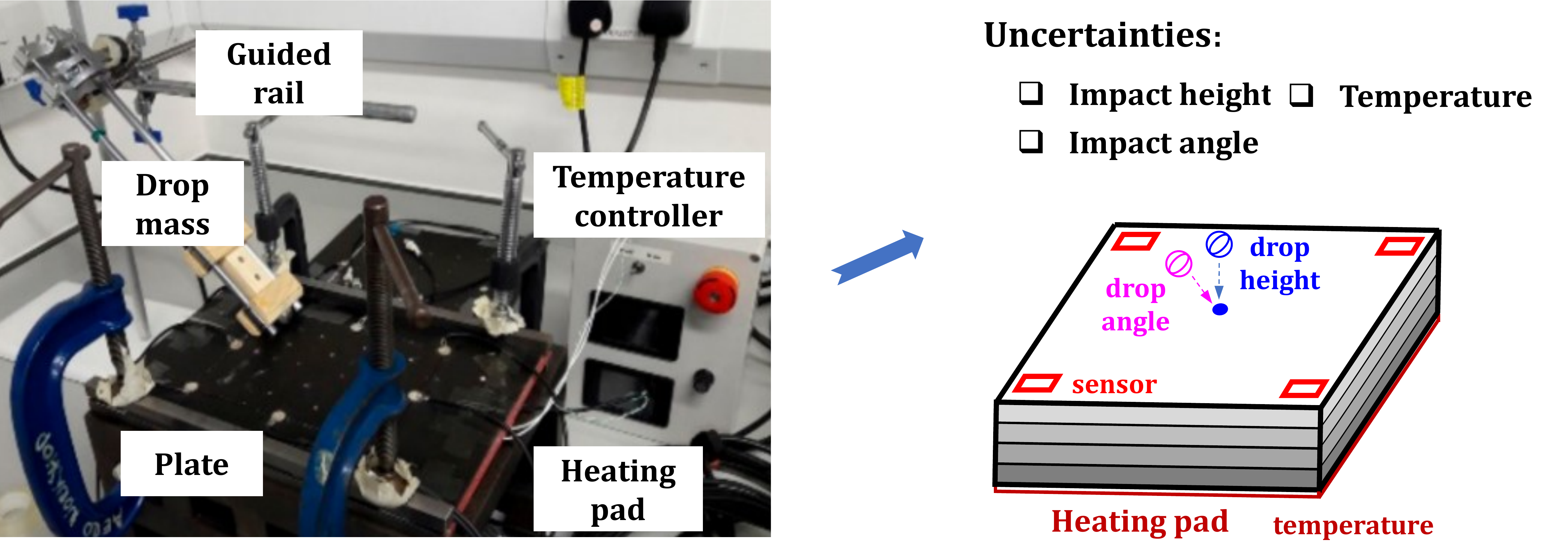}
	\caption{Experimental setup for guided drop mass impact testing, considering uncertainties in impact height, angle and temperature. }
	\label{FIG: exp guided drop mass}
\end{figure}

To investigate the effects of impact angle, height, and environmental temperature on the composite plate and the accuracy of impact localisation, experiments were conducted using a guided drop mass system. As shown in \cref{FIG: exp guided drop mass}, the plate was clamped along its longitudinal edges, while the drop mass was released along a guided rail, providing precise control over both impact angle and height. This setup ensured high repeatability and minimized variability caused by unintended deviations, enabling a systematic examination of how these parameters influence the dynamic response of the plate. In addition, a heating pad was installed on the underside of the plate, facilitating controlled temperature adjustments. During the guided drop mass testing, six sensors, numbered 1 to 6 in \cref{FIG: plate_illustration}, were employed to capture the impact-induced responses. These sensors were connected to a PXI-5105 oscilloscope operating at a sampling frequency of 2 MHz. 

Four different impact cases were performed on the composite panel using the guided drop mass, as detailed in \cref{TB: exp impact cases guided drop mass}. The reference (REF) impacts were conducted perpendicularly at 35 evenly distributed locations, numbered 1-35 in \cref{FIG: plate_illustration}, using a 100 g impact mass dropped from a height of 1 cm at room temperature (24 $^\circ$C). In contrast, three additional cases introduced variations in height, angle and temperature: impact height increased to 2 cm (HEI), impact angle adjusted to 45 $^\circ$ (ANG), and temperature elevated to 70 $^\circ$C (TEM). To mitigate the influence of experimental randomness, each case was repeated two times, ensuring robust and reliable data collection.

\begin{table}[ht]
\centering
\caption{Impact testing conducted using guided drop mass accounting uncertainties of impact height, angle and temperature.}\label{TB: exp impact cases guided drop mass}
\begin{tabular}{cccccc}
\toprule
Impact case  & Temperature ($^\circ$C) & Mass (g) & Height (cm) & Angle & Number of impacts\\
\midrule
REF & 24 & 100 & 2.5 & $90^\circ$ & 35 * 2 Rep \\
HEI & 24 & 100 & 5 & $90^\circ$ & 35 * 2 Rep \\
ANG & 24 & 100 & 2.5 & $45^\circ$ & 35 * 2 Rep \\
TEM & 70 & 100 & 2.5 & $90^\circ$ & 35 * 4 Rep \\
\bottomrule
\end{tabular}\\
    \begin{tablenotes}    
        \footnotesize               
        \item[1] \hspace{2mm} *Note: REF = reference, HEI = height, ANG = angle, TEM = temperature, Rep = repetitions.
  \end{tablenotes}            
\end{table}

The data collected under temperature variations (TEM case) was generated by Dr. Aldyandra Hami Seno\href{https://orcid.org/0000-0001-9945-5299}{\includegraphics[scale=0.04]{orcidicon.pdf}}, a former Ph.D. student in the research group. Detailed information on experimental data collection under temperature variations can be found in \cite{seno_passive_2019, seno_uncertainty_2021}.

\section{TDOA extraction under uncertainties} \label{section TDOA extraction}
A reliable TDOA extraction method capable of consistently estimating TDOA under various uncertainties is critical for robust, wave propagation-based, data-driven impact identification. While the Normalized Smoothed Envelope Threshold (NSET) method \cite{seno_passive_2019, hami_seno_multifidelity_2023} has demonstrated high accuracy and consistency in TDOA extraction for fixed impact scenarios with specific parameters \cite{xiao_time_2024}, its robustness under significant uncertainties—such as variations in impact mass and impactor—requires experimental validation.

\subsection{TDOA uncertainties due to impact mass and impactor} \label{TDOA uncertainties due to impact mass and impactor}
Since impact mass and impactor material significantly affect impact responses \cite{olsson_mass_2000}, two distinct impact scenarios were considered: large-mass drop tower impacts (DT) and small-mass hammer impacts (HA), both conducted at location D, as shown in \cref{FIG: plate_illustration}.

\begin{figure}[htb] 
	\centering
		\includegraphics[width=1\columnwidth]{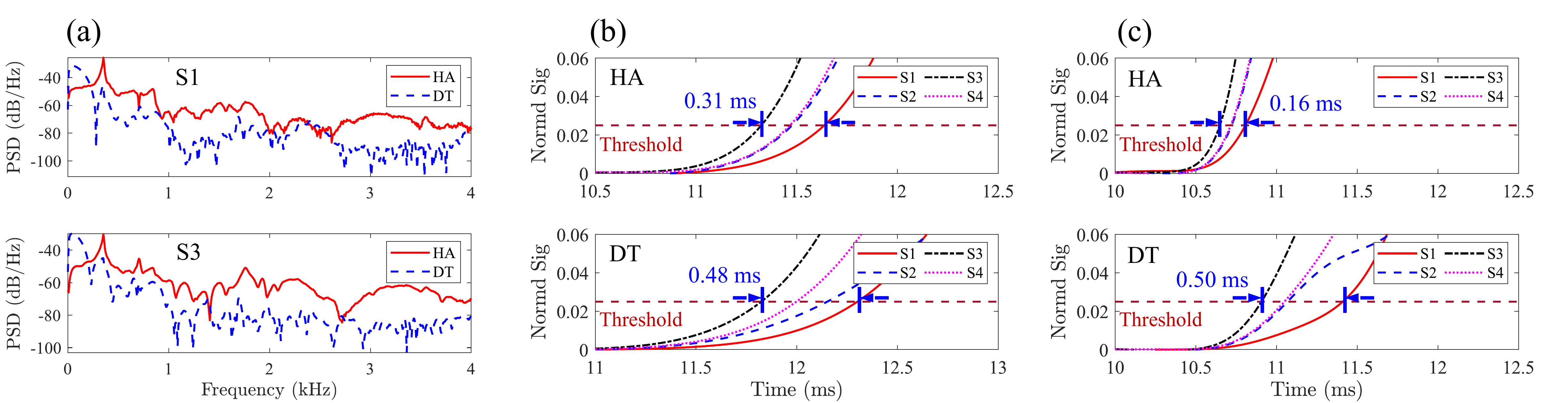}
	\caption{TDOA extracted by NSET method for HA and DT impacts at location D with frequencies: (a) PSD, (b) 1 kHz, (c) 2 kHz.}
	\label{FIG: NSET TDOA difference}
\end{figure}

\cref{FIG: NSET TDOA difference}(a) compares the power spectral density (PSD) of normalised impact sensor signals recorded by sensors 1 and 3 for DT and HA impacts. The results are consistent with the dynamics of the impacts. Large-mass DT impacts are dominated by low-frequency components, with PSD values decreasing sharply as the frequency increases. Most of the signal energy is concentrated below 0.5 kHz. In contrast, small-mass HA impacts exhibit a broader range of dominant frequencies (below 2 kHz) with a more gradual decline in PSD.

These spectral differences have a pronounced effect on TDOA. The TDOA values extracted using the NSET method for HA and DT impacts at location D at frequencies of 0.5 kHz and 1 kHz, with a threshold of 2.5$\%$ of the maximum signal amplitude, are presented in \cref{FIG: NSET TDOA difference}(b)-(c). The results indicate that the TDOA order remains consistent across variations in impact mass, with waves consistently arriving first at sensor 3 and last at sensor 1. However, the magnitude of the TDOA varies significantly, reflecting challenges in maintaining precise distance consistency.

At 0.5 kHz, the maximum TDOA between the first-arriving and last-arriving sensors for large-mass DT impacts is 0.48 ms, slightly exceeding the 0.31 ms observed for HA impacts. At 1 kHz, however, the maximum TDOA for DT impacts increases anomalously to 0.50 ms, more than three times the 0.16 ms observed for HA impacts. This finding contradicts the principle of flexural wave dispersion, where higher frequencies correspond to higher wave velocities, resulting in shorter TDOA intervals. While this principle holds for HA impacts, it is violated in the case of DT impacts increasing from 0.5 kHz to 1 kHz.

The substantial maximum TDOA observed for DT impacts at 1 kHz and the deviation from expected dispersion behavior are primarily attributed to differences in the frequency components dominating the initial segments of the filtered envelopes. The small-mass HA impacts, with their broader range of dominant frequencies, adhere to the expected dispersion relationship below 2 kHz. In contrast, for DT impacts, the signal component at 1 kHz is negligible, leading to TDOA estimates being derived from lower dominant frequencies. Consequently, TDOA discrepancies become pronounced due to wave dispersion effects. These findings underscore the challenges posed by diverse frequency distributions induced by varying impact scenarios. The observed TDOA differences highlight the necessity for impact localisation methods that account for frequency-dependent variations. Such methods are crucial for enhancing the robustness and consistency of TDOA extraction, particularly in practical applications where impact conditions are subject to significant variability.

\subsection{TDOA uncertainties due to impact height, angle and temperature}
\begin{figure}[htb] 
	\centering
		\includegraphics[width=0.8\columnwidth]{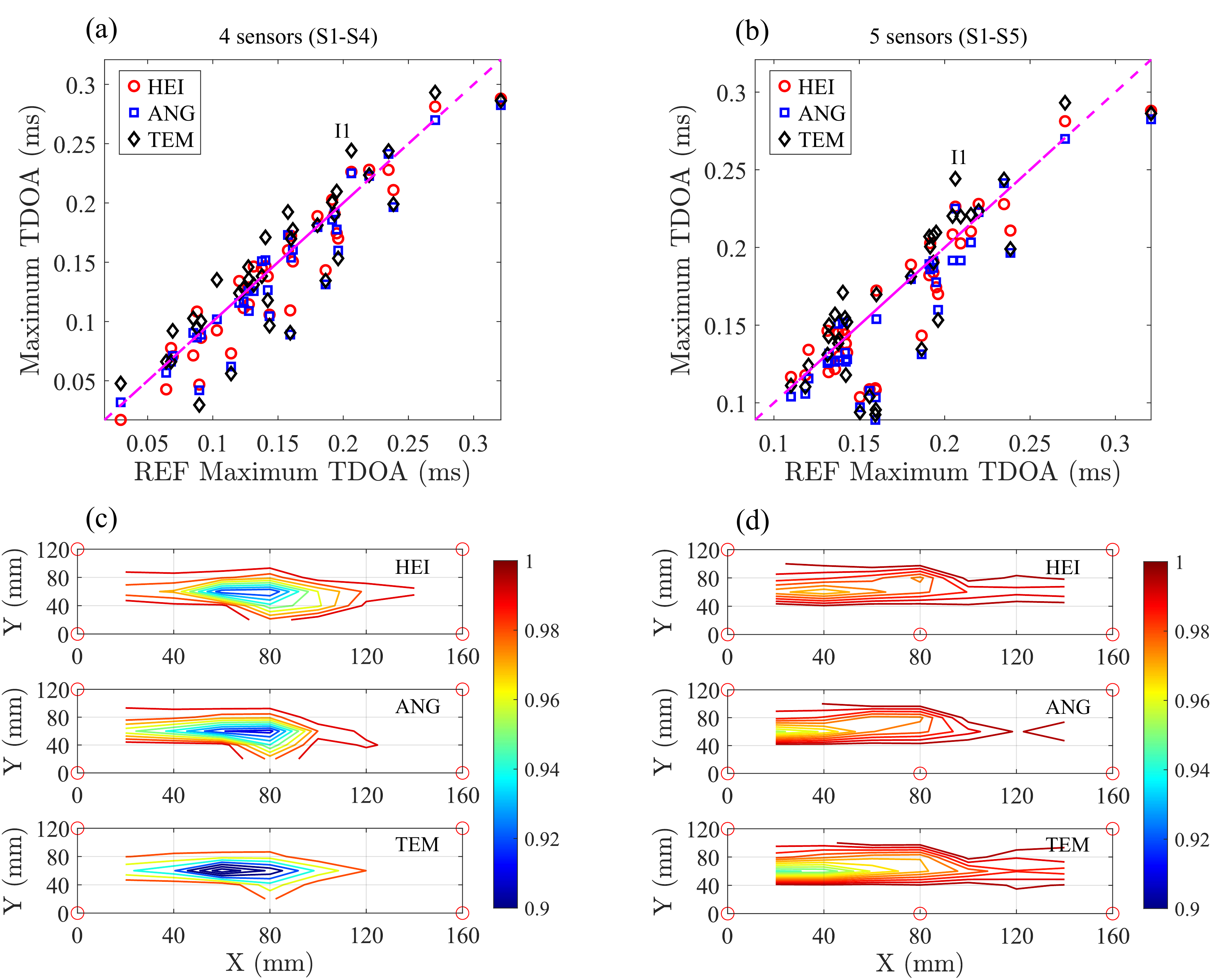}
	\caption{TDOA values extracted using the NSET method for four cases of drop mass impact testing: (a)-(b) Maximum TDOA differences between REF impacts and the HEI, ANG, and TEM cases, with 4 and 5 sensors, respectively; (c)-(d) Cosine similarity of TDOA between REF impacts and the HEI, ANG, and TEM cases, with 4 and 5 sensors, respectively. }
	\label{FIG: TDOA uncertainties}
\end{figure}

The four guided drop mass impact cases were used to examine TDOA uncertainties arising from variations in impact height, angle, and temperature. \cref{FIG: TDOA uncertainties} presents the TDOA values extracted using the NSET method for these scenarios, with both 4 and 5 sensors, respectively. As shown in \cref{FIG: TDOA uncertainties}(a)-(b), the maximum TDOA differences across all 35 impact locations, induced by changes in impact height (HEI), impact angle (ANG), and a 46 $^\circ$C temperature increase (TEM), consistently remain within approximately $\pm 0.05$ ms of the reference (REF) impacts, regardless of the number of sensors used. For impacts with a maximum TDOA below 0.1 ms using 4 sensors, the differences between impact cases are more pronounced. However, when an additional sensor (sensor 5) is introduced between sensor 1 and sensor 2, the lower bound of the maximum TDOA increases (exceeding 0.1 ms), indicating a reduced singularity in TDOA, as discussed in \cref{Kernel design}.

This reduction in singularity is also reflected in the cosine similarity between the REF impacts and the HEI, ANG, and TEM cases, which measures the order similarity of TDOAs. As shown in \cref{FIG: TDOA uncertainties}(c)-(d), the contour plots of cosine similarity across the plate indicate that the cosine similarity consistently exceeds 0.9 and 0.95 for 4 and 5 sensors, respectively. Impacts located closer to the sensors, where variations in impact-to-sensor distances are larger and TDOA values are higher, exhibit greater cosine similarity. Conversely, in the central region of the plate, where impact-to-sensor distances are nearly identical and TDOA values are minimal, the cosine similarity decreases slightly. This behavior aligns with the proposed analysis regarding the influence of singularity point on TDOA consistency.

The effects of environmental temperature on TDOA differ significantly from those of impact height and angle. From a thermal expansion perspective, an increase in temperature reduces bending stiffness of the structure, which in turn decreases wave propagation velocities and increases TDOA values. However, as shown in \cref{FIG: TDOA uncertainties}(a)-(b), this trend is not consistently observed across all impact locations. Approximately two-thirds of the impacts follow this trend, showing the highest TDOA values in the TEM case. This can be attributed to two primary factors: First, for impacts near the singularity point, characterized by smaller TDOA values, the TDOA measurements become more sensitive to signal noise and errors in the TDOA extraction method. Second, the threshold used to extract TDOA in the NSET method is set to 0.25$\%$ of the signal peak, which varies for each impact. This adaptive threshold can result in either overestimation or underestimation of the TDOA, depending on the amplitude of the signal peak. 

Similar to the findings for DT impacts, impact localisation solely based on distance similarity of TDOA values proves insufficient to address uncertainties related to impact mass and temperature. However, the high similarity in TDOA order, as denoted by the cosine similarity in \cref{FIG: TDOA uncertainties}(b), suggest a promising avenue for improving localisation accuracy and reliability. By employing a composite input kernel that combines distance similarity (RBF kernel) with order similarity (COS kernel), it is anticipated that the robustness of the localisation process under uncertainties can be significantly enhanced.

\section{Impact localisation under environmental and operational uncertainties}  \label{section: Impact Localisation}
This section presents the experimental validation and analysis of impact localisation on aircraft structures using various kernels and Bayesian kernel fusion under environmental and operational uncertainties. The study also explores the role of data pre-processing techniques, specifically two standardisation methods: sample standardisation (SS) and feature standardisation (FS). These approaches aim to address the challenges of localisation under uncertain conditions and can be mathematically expressed using the training input matrix $X$, an $N$ by $M$ matrix with $N$ samples and $M$ features, as follows:
\begin{equation}
\begin{array}{lcl} \label{EQU: standardization approaches}
    X_{ss}(i,:) = \frac{X(i,:)}{\left\| X(i,:)  \right\| }, \; 
    X_{fs}(:,j) =  \frac{X(:,j) - \mean[X(:,j)] }{\std[X(:,j)]}.
\end{array}
\end{equation}  

Here, $X_{ss}$ represents the sample-standardised data, where each row (sample) of $X$ is scaled to a unit vector by dividing by its Euclidean norm, effectively normalising the magnitude of the TDOA vector for each impact. $X_{fs}$, on the other hand, represents the feature-standardised data, where each column (feature) is transformed to have zero mean and unit variance, ensuring consistency across sensors. These standardisation methods are evaluated for their effectiveness in enhancing impact localisation performance under varying uncertainties.

\subsection{Comparative analysis of kernels and Bayesian kernel fusion for impact localisation} \label{subsection: kernels and Bayesian kernel fusion}
To investigate the effects of different input kernels in GPR on impact localisation, the TEM impact case, which exhibits the most significant deviation in TDOA values from the reference (REF) case due to temperature changes, was selected. Three input kernels were examined: cosine similarity (COS), radial basis function (RBF), and a composite kernel (COMP), which multiplicatively combines COS and RBF. Furthermore, kernel fusion was applied using Bayesian Model Averaging (BMA) to integrate predictions from multiple kernels. In the GPR model, the input TDOA vector for each impact was normalised to a unit vector using sample standardization, achieved by dividing each vector by its modulus. The output of the GPR model, the impact location coordinates in millimeters, was directly used without further transformation. 

\begin{figure}[htb] 
	\centering
		\includegraphics[width=0.8\columnwidth]{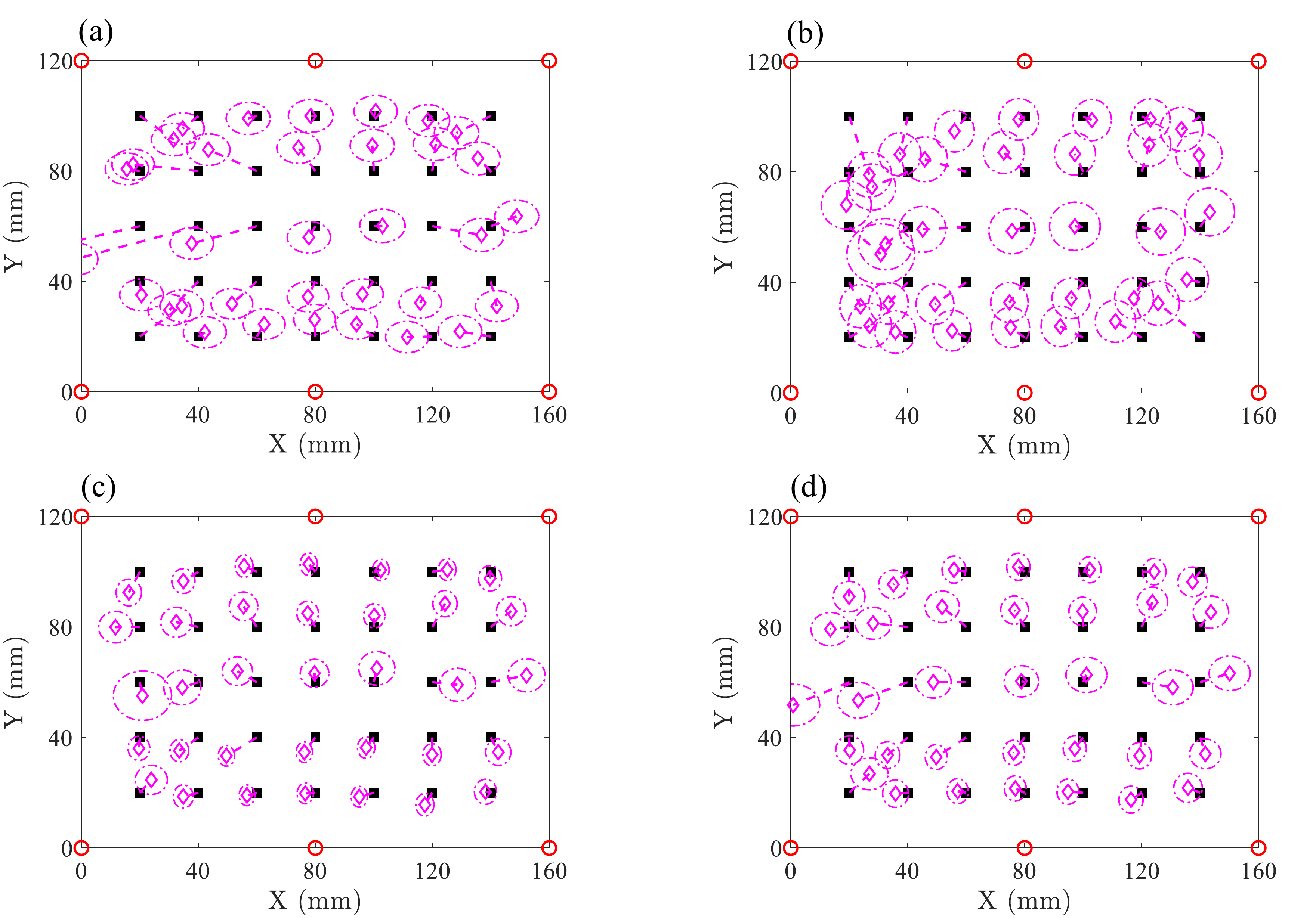}
	\caption{Multitask impact localisation for TEM impacts based on REF impacts using different input kernels: (a) COS kernel, (b) RBF kernel, (c) COMP kernel, (d) BMA fusion. }
	\label{FIG: fig_loc_TEM_illustration}
\end{figure}

\cref{FIG: fig_loc_TEM_illustration} illustrates the localisation results for TEM impacts based on REF impacts, leveraging these kernels and BMA fusion with six passive sensors. In this figure, red hollow circles represent sensor locations along the X and Y axes, black solid squares denote actual impact locations, and magenta diamonds indicate estimated impact positions. Localisation uncertainties (standard deviation) are visualised as dashed magenta circles around each estimate, providing insight into the performance of each kernel under temperature-induced uncertainties.

The results in \cref{FIG: fig_loc_TEM_illustration}(a)-(c) reveal distinct advantages and limitations of each kernel. The COS kernel demonstrates accurate localisation near sensors but performs poorly near the singularity region (e.g., along the line $Y$ = 60 mm), where cosine similarities with small TDOA values are more susceptible to noise. Conversely, the RBF kernel excels in the singularity region due to its reliance on TDOA amplitude, but shows reduced accuracy near sensors. By combining the strengths of both kernels, the COMP kernel achieves superior localisation accuracy across the entire plate. As shown in \cref{FIG: fig_loc_TEM_illustration}(d), the localisation results obtained through BMA fusion closely resemble those of the COMP kernel, indicating that BMA effectively identifies and assigns higher weights to the COMP kernel compared to COS and RBF. 

\begin{figure}[htb] 
	\centering
		\includegraphics[width=1\columnwidth]{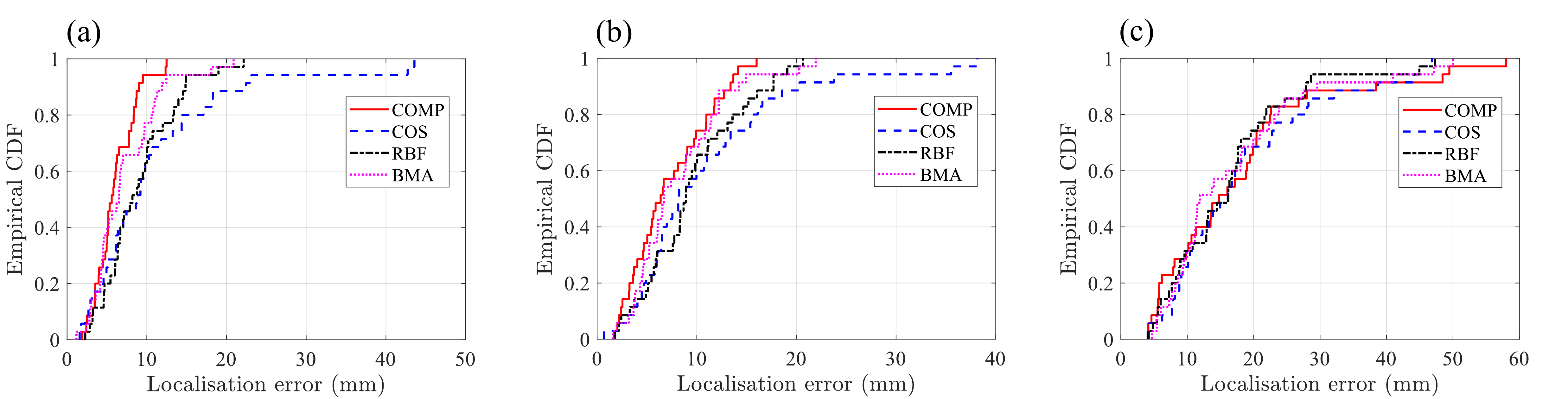}
	\caption{Empirical CDF of localisation error for TEM impacts based on REF impacts, under variations in sensors: (a) 6 sensors (S1-S6), (b) 5 sensors (S1-S5), (c) 4 sensors (S1-S4). }
	\label{FIG: fig_loc_TEM_varysensors}
\end{figure}

The empirical cumulative distribution function (CDF) of localisation errors for 35 TEM impacts, using six sensors, is presented in \cref{FIG: fig_loc_TEM_varysensors}(a). The COMP kernel outperforms both the COS and RBF kernels, with the CDF rising rapidly to 1 and a maximum localisation error of 12.4 mm, compared to 43.5 mm and 22.1 mm for the COS and RBF kernels, respectively. The mean localisation errors further highlight this improvement: the COMP kernel achieves a mean error of 6.1 mm, compared to 11.1 mm for COS and 9.0 mm for RBF. The BMA fusion strategy demonstrates performance similar to the COMP kernel, with a maximum localisation error of 20.9 mm and a mean error of 7.2 mm. These findings validate the COMP kernel’s ability to enhance localisation accuracy by combining order similarity (COS) and distance similarity (RBF), and underscore the effectiveness of BMA in improving the robustness of predictions by leveraging multiple kernels.

To evaluate the robustness of BMA under varying sensor configurations, \cref{FIG: fig_loc_TEM_varysensors}(b)-(c) presents the empirical CDFs of localisation errors for TEM impacts using five and four sensors, respectively. Together with \cref{FIG: fig_loc_TEM_varysensors}(a), which shows results for six sensors, the analysis demonstrates that BMA consistently outperforms the COS and RBF kernels, regardless of the number of sensors. Even with only four sensors, the BMA achieves a mean localisation error of 16.2 mm, slightly better than the 17.8 mm and 16.3 mm achieved by the COS and RBF kernels, respectively.

The results further highlight the influence of sensor quantity on localisation accuracy. As the number of sensors decreases from six to five and then to four, localisation accuracy decreases correspondingly. For example, the mean localisation error for BMA increases from 7.2 mm with six sensors to 8.2 mm with five sensors and 16.2 mm with four sensors. This trend underscores the importance of sensor density in achieving optimal localisation performance while showcasing the robustness of BMA in adapting to reduced sensor configurations. The combination of kernel design and fusion strategies ensures reliable and accurate impact localisation, even under challenging environmental and operational uncertainties.




\subsection{Data preprocessing: sample standardization (SS) vs. feature standardisation (FS)}
In machine learning, feature standardisation (FS) is a widely used technique to transform data, ensuring that individual features (e.g., the TDOA of a sensor) are scaled to a similar range, which typically enhances model performance. As defined in \cref{EQU: standardization approaches}, this approach involves applying a constant shift and scaling factor independently to each feature (TDOA from a sensor). While FS maintains the relative magnitudes of TDOAs across impacts, it disrupts the internal order of components within the input TDOA vector due to the non-uniform shifts and scaling applied across different sensors.

In contrast, sample standardisation (SS) normalises the entire TDOA vector for each impact by dividing it by its modulus, transforming it into a unit vector. Unlike FS, SS preserves the internal order of components within each TDOA vector while ensuring that the magnitude of every TDOA vector is standardised to unity. This makes SS particularly effective in managing scale variations in input TDOAs caused by environmental and operational uncertainties, ensuring that the inputs to the GPR model are consistent and appropriately scaled for accurate kernel-based modelling.

\begin{figure}[htb] 
	\centering
		\includegraphics[width=0.8\columnwidth]{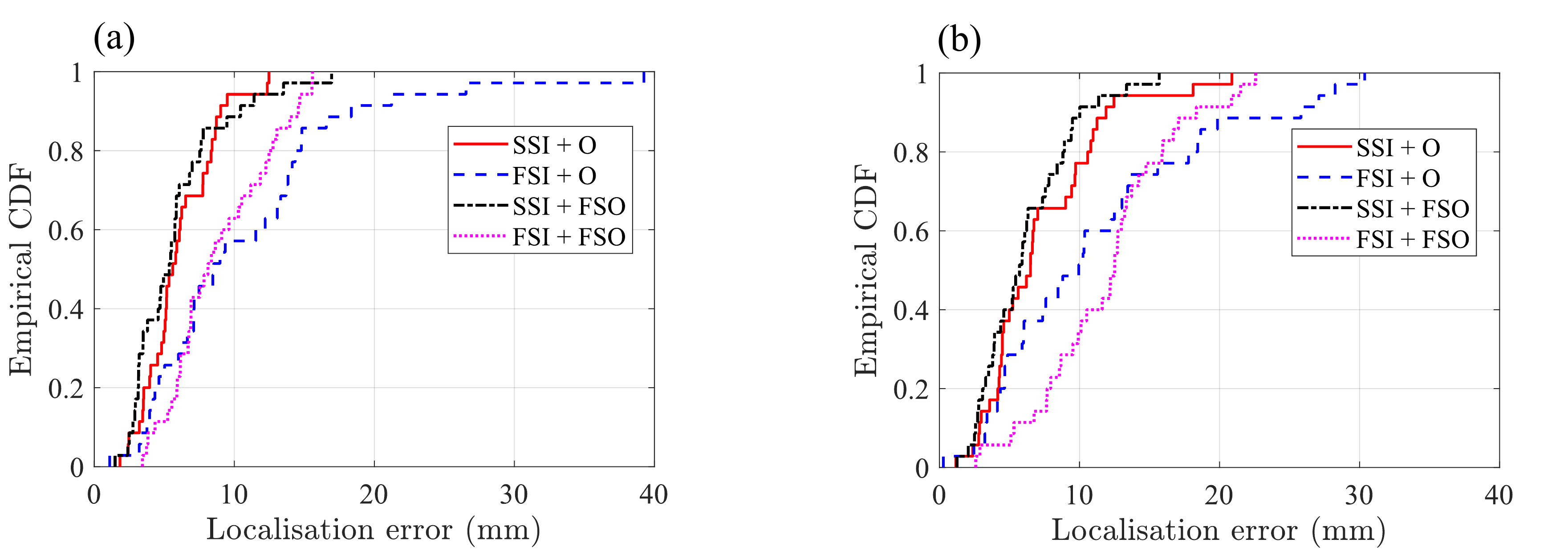}
	\caption{Empirical CDF of localisation error for TEM impacts based on REF impacts, comparing different data preprocessing approaches with: (a) composite (COMP) kernel, (b) Bayesian Model Averaging (BMA) fusion. }
	\label{FIG: fig_loc_TEM_varynor}
\end{figure}

As demonstrated in \cref{subsection: kernels and Bayesian kernel fusion}, normalising the input TDOA vectors using sample standardisation achieved high accuracy in impact localisation. To further evaluate the performance of different preprocessing methods, a comparative study of sample standardisation (SS) and feature standardisation (FS) applied to the input TDOAs was conducted. \cref{FIG: fig_loc_TEM_varynor} presents the empirical CDF of localisation errors for TEM impacts, comparing these preprocessing approaches with the composite (COMP) kernel and Bayesian Model Averaging (BMA) fusion. In the legend, 'I' and 'O' represent the input TDOA in milliseconds and the output location coordinates in millimeters, respectively. 'SSI' refers to sample standardisation applied to inputs, while 'FSI' and 'FSO' denote feature standardisation applied to inputs and outputs, respectively.

The results indicate that for both the COMP kernel and BMA fusion, sample standardisation of inputs (SSI) significantly outperforms feature standardisation of inputs (FSI) in achieving accurate TDOA-based impact localisation. For instance, the mean localisation error using SSI with the COMP kernel and BMA fusion ('SSI + O') is 6.1 mm and 7.2 mm, respectively, compared to 10.7 mm and 11.2 mm for 'FSI + O.' This superior performance of SSI can be attributed to its ability to preserve the relative order within TDOA vectors and to effectively handle scale variations. 

The underperformance of FSI stems from its disruption of order similarity, which is critical for the COS kernel, and its inability to address scale variations, a key aspect for the RBF kernel. By retaining order similarity and normalising TDOA magnitudes consistently, SSI ensures compatibility with both order- and distance-based kernels, leading to improved impact localisation accuracy. 

Additionally, the effects of feature standardisation on output coordinates were investigated. As shown in \cref{FIG: fig_loc_TEM_varynor}, applying feature standardisation to output coordinates (FSO) slightly improves localisation accuracy when input TDOAs are normalised by either sample standardisation (SS) or feature standardisation (FS). This improvement is particularly evident in the reduction of maximum localisation errors, as shown in the comparisons between 'SSI + O' and 'SSI + FSO', as well as 'FSI + O' and 'FSI + FSO'.

After output standardisation, the BMA fusion method achieves a mean localisation error of 6.1 mm and a maximum error of 15.7 mm for 'SSI + FSO', compared to 7.2 mm and 20.9 mm for 'SSI + O'. This improvement is attributed to enhanced numerical stability, as FSO mitigates kernel matrix ill-conditioning and accelerates the convergence of the marginal log-likelihood during model training. Furthermore, BMA fusion assigns greater weights to more accurate kernels, thereby increasing the robustness of predictions. Considering numerical stability, model interpretability, and overall performance, feature standardisation of output coordinates is recommended as a complementary preprocessing step.


\subsection{Impact localisation under various uncertainties}
\begin{figure}[htb] 
	\centering
		\includegraphics[width=0.8\columnwidth]{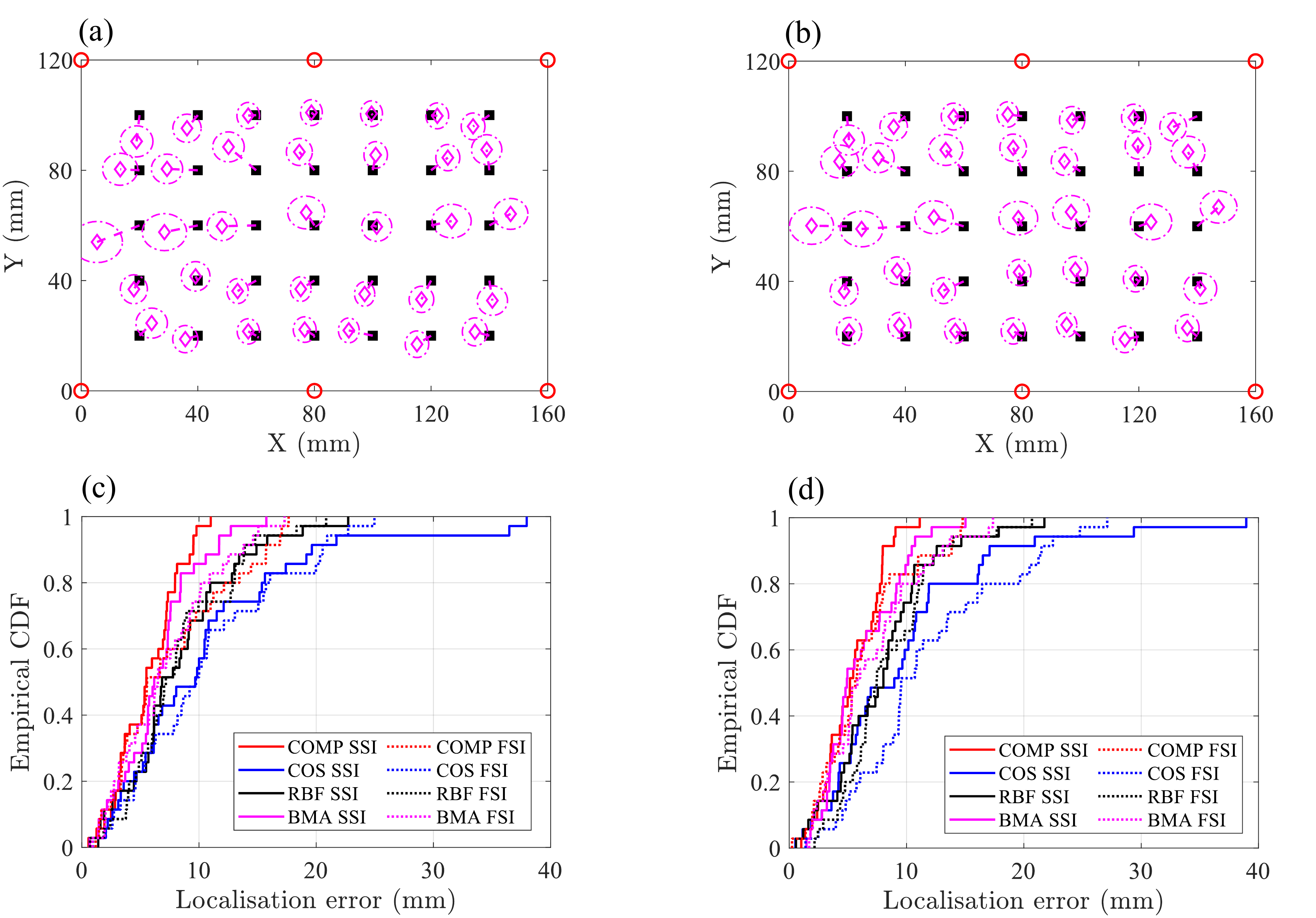}
	\caption{Localisation of ANG and HEI impacts based on REF impacts: (a) localisation illustration of ANG impacts using BMA fusion with SDI, (b) localisation illustration of HEI impacts using BMA fusion with SDI, (c) empirical CDF of localisation errors for ANG impacts, comparing SDI and FSI, (d) empirical CDF of localisation errors for HEI impacts, comparing SDI and FSI. }
	\label{FIG: fig_loc_ANG_HEI_varynor}
\end{figure}

To evaluate the performance of impact localisation under uncertainties in impact height and angle, the ANG and HEI impacts were localised using the reference (REF) impacts recorded with six sensors. As shown in \cref{FIG: fig_loc_ANG_HEI_varynor}(a)-(b), the proposed BMA fusion demonstrated high accuracy in localising these impact cases under uncertain conditions. 

For the ANG case, the mean and maximum localisation errors were 6.5 mm and 15.7 mm, respectively, with corresponding standard deviations (SDs) of 7.1 mm and 11.5 mm. Similarly, for the HEI case, the mean and maximum localisation errors were 6.1 mm and 15.0 mm, with SDs of 7.3 mm and 10.1 mm. These results highlight the robustness of the BMA fusion approach in handling uncertainties in impact conditions while maintaining precise localisation performance. 

The normalisation approaches for input TDOAs, specifically sample standardisation (SSI) and feature standardisation (FSI), were also evaluated for the ANG and HEI impacts. The empirical CDFs of localisation errors, as illustrated in \cref{FIG: fig_loc_ANG_HEI_varynor}(c)-(d), provide a comparative analysis. In these figures, solid lines represent SSI, while dashed lines represent FSI. Different colors are used to distinguish the various kernels and the BMA fusion.

Across all kernels and the BMA fusion, the solid lines (SSI) consistently appear slightly to the left of the dashed lines (FSI) for the same kernel, indicating improved localisation accuracy with SSI. This demonstrates the effectiveness of SSI in impact localisation, even though FSI also achieves high accuracy. Additionally, the superior performance of the COMP kernel is reaffirmed, showing better accuracy compared to the COS and RBF kernels. This enhancement is further reflected in the BMA fusion, which achieves higher localisation accuracy than the individual COS and RBF kernels, reinforcing the benefits of kernel design and fusion techniques for robust impact localisation.

\subsection{Impact localisation under varying boundary condition and impact mass}
TDOA-based impact localisation methods are expected to remain effective regardless of structural boundary conditions, owing to the independence of the impact wavefront from these conditions. To validate this hypothesis, drop tower (DT) impacts were localised using guided drop mass (REF) impacts as references. As shown in \cref{Experimental setup}, DT impacts were conducted on a plate simply supported along its longitudinal edges, whereas REF impacts were conducted on the same plate fully clamped along the longitudinal edges. Additionally, as illustrated in \cref{TDOA uncertainties due to impact mass and impactor}, the significant difference in impact mass (5500 g for the DT impact vs. 100 g for the REF impacts) resulted in distinctly different spectral distributions of the impact responses. This experiment also aimed to evaluate the performance of impact localisation under uncertainties associated with impact mass.

\begin{figure}[htb] 
	\centering
		\includegraphics[width=0.8\columnwidth]{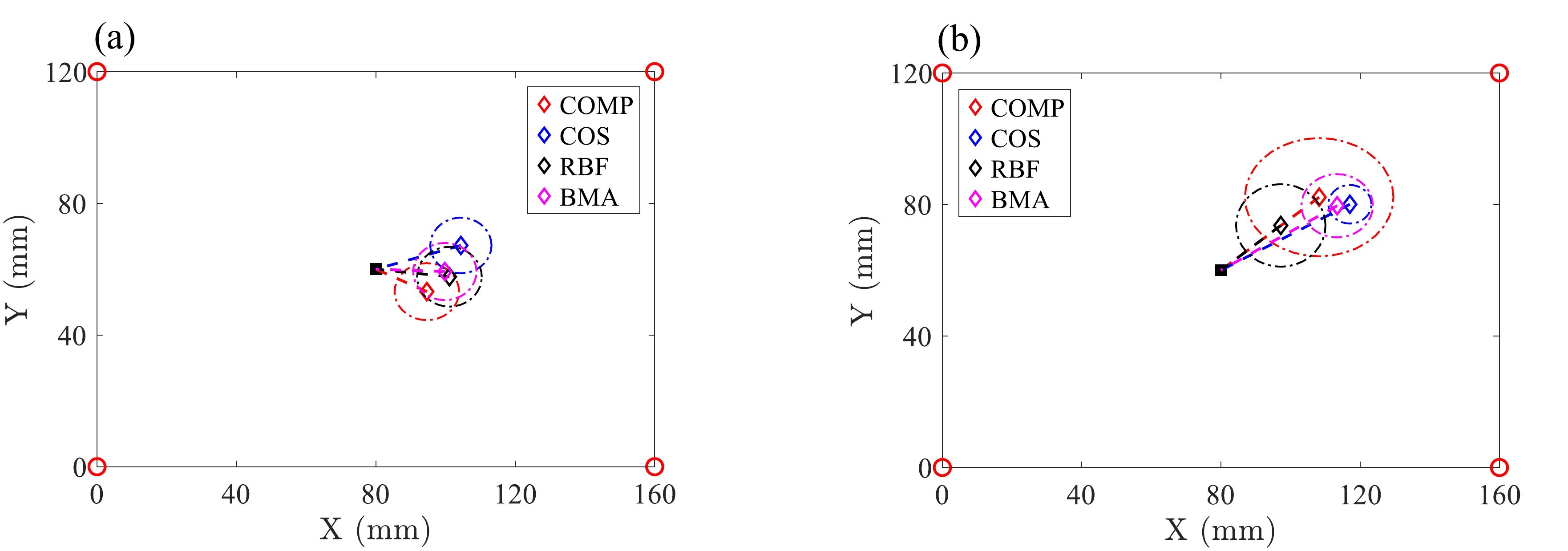}
	\caption{Localisation of DT impact based on REF impacts by kernel design and BMA fusion: (a) sample standardisation (SSI), (b) sample standardisation (FSI).  }
	\label{FIG: fig_loc_DT_varynor}
\end{figure}

\cref{FIG: fig_loc_DT_varynor} presents the localisation of DT impacts based on REF impacts using various kernel designs and BMA fusion, comparing the performance of sample standardisation (SSI) and feature standardisation (FSI). Due to the substantially larger TDOA magnitudes of the DT impacts compared to the REF impacts, SSI consistently achieved higher localisation accuracy than FSI across all three kernels and BMA fusion. The localisation errors with SSI for the COMP, COS, RBF kernels, and BMA were 16.1 mm, 25.4 mm, 21.2 mm, and 19.7 mm, respectively. In contrast, the corresponding errors for FSI were significantly higher: 35.9 mm, 42.1 mm, 21.9 mm, and 38.7 mm, respectively.

Moreover, with SSI, the RBF and COMP kernels demonstrated smaller SDs, indicating higher confidence in localisation, compared to FSI. Consequently, under SSI, the BMA fusion assigned nearly equal weights to the COMP, COS, and RBF kernels. However, with FSI, the BMA was dominated by the COS kernel, which had the lowest SD but the highest localisation error, highlighting a limitation of relying solely on SD as an indicator of localisation accuracy.

These findings underscore the importance of sample standardisation (SSI) for accurate and reliable impact localisation under uncertainties in impact mass. By effectively managing scale variations in TDOAs, SSI ensures compatibility across kernels, leading to improved localisation performance and robust fusion outcomes.

\subsection{Influence of variability in reference impact data set}
\begin{figure}[htb] 
	\centering
		\includegraphics[width=0.8\columnwidth]{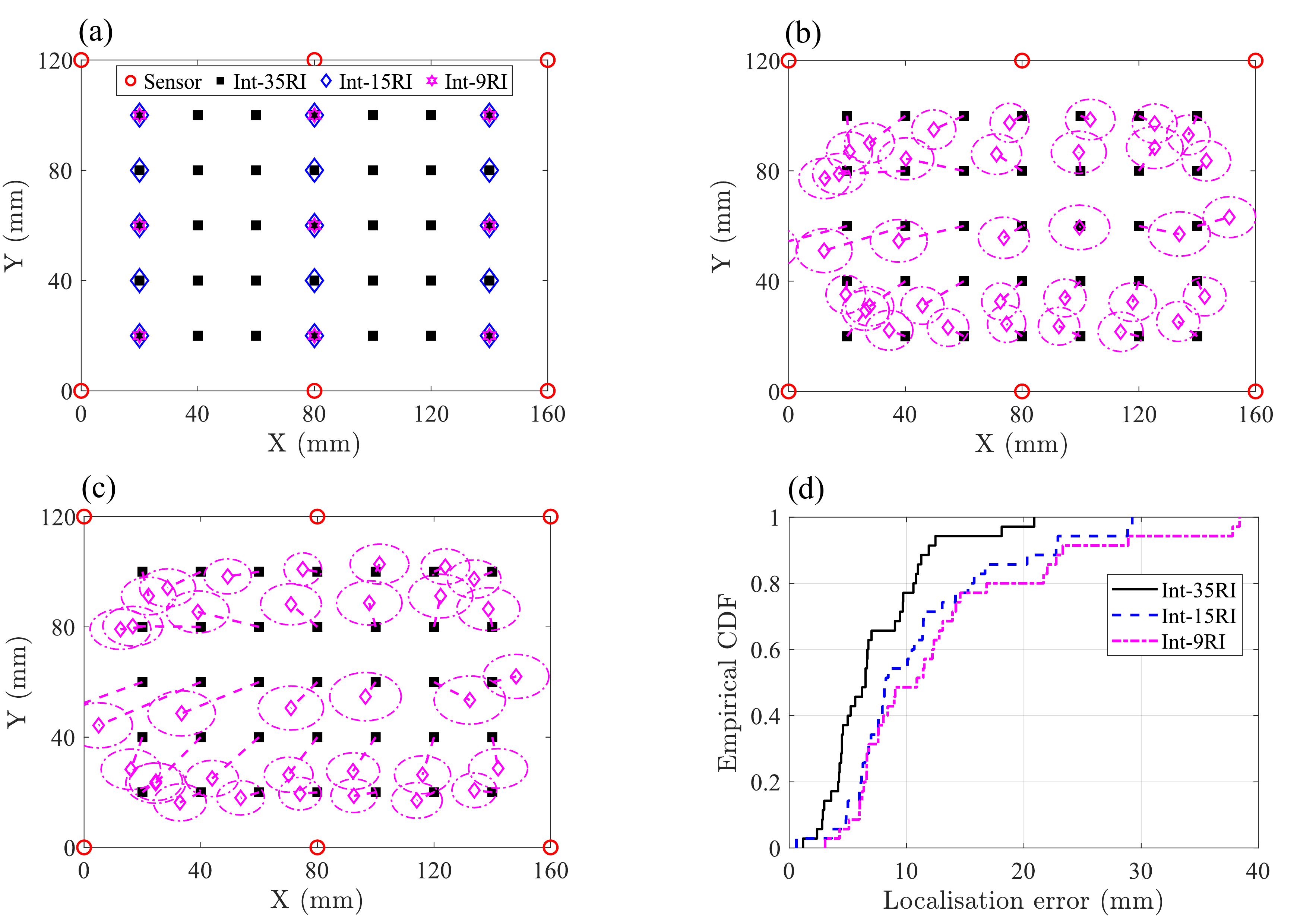}
	\caption{Interpolated localisation of TEM impacts using different sizes of reference impact (RI) sets with kernel design and BMA fusion: (a) illustration of the three RI sets, (b) impact localisation with 15 RI using BMA fusion, (c) impact localisation with 9 RI using BMA fusion, (d) empirical CDF of impact localisation errors for the three RI sets using BMA fusion.}
	\label{FIG: fig_loc_diff_train_interpolation}
\end{figure}

The above analysis of impact localisation methodology in this study was developed and validated using 35 reference impacts (RI) densely distributed on a composite plate in a 20 mm by 20 mm grid. To investigate the effect of the number of reference impacts on localisation performance, two additional subsets were created by systematically reducing the number of RI to 15 and 9, respectively, from the original set of 35. \cref{FIG: fig_loc_diff_train_interpolation}(a) illustrates the three reference impact configurations: black solid squares represent the full set of 35 RI, blue diamonds denote the subset with 15 RI, and magenta hexagrams indicate the subset with 9 RIs.

The subset with 15 RI was constructed by selecting the first, fourth, and seventh columns from the original 35 RI set, while the subset with 9 RI was derived by further excluding impacts located in the second and fourth rows, making it a subset of the 15 RI configuration. Compared to the 9 RI subset, the 15 RI configuration is expected to improve localisation accuracy in the regions of the second and fourth rows due to the increased density of reference impacts in these areas.

Importantly, all three reference sets maintain the same rectangular coverage area, ensuring that the localisation of all impact cases remains an interpolation (referred to as 'Int' in the figure legend) problem. This setup offers valuable insights into the trade-offs between reference impact density and localisation accuracy, specifically under interpolation conditions.

The impact localisation results for TEM impacts using 15 reference impacts (RI) and 9 RI are illustrated in \cref{FIG: fig_loc_diff_train_interpolation}(b) and (c), respectively. When combined with the localisation results of the complete set of 35 RIs, shown in \cref{FIG: fig_loc_TEM_illustration}(d), and the empirical CDF of localisation errors for the three RI sets, depicted in \cref{FIG: fig_loc_diff_train_interpolation}(d), it becomes evident that increasing the number of reference impacts improves localisation accuracy.

As expected, the set of 15 RI demonstrates enhanced localisation accuracy in the second and fourth rows, attributed to the increased data density in these regions compared to the set of 9 RI. Specifically, the 15 RI configuration achieves mean and maximum localisation errors of 10.9 mm and 29.3 mm, respectively, compared to 12.9 mm and 38.4 mm for the 9 RI configuration. Moreover, the complete set of 35 RI significantly improves overall localisation accuracy, with mean and maximum errors reduced to 7.2 mm and 20.9 mm, respectively. 

The mean localisation errors of approximately 10 mm across all three RI sets demonstrate the high accuracy of the proposed impact localisation methods based on kernel design and BMA fusion. As detailed in \cref{subsection: kernels and Bayesian kernel fusion}, this accuracy is attributed to the COMP kernel, which combines the strengths of the COS and RBF kernels, and the effectiveness of BMA fusion, which assigns higher weights to more accurate predictions.

\begin{figure}[htb] 
	\centering
		\includegraphics[width=1\columnwidth]{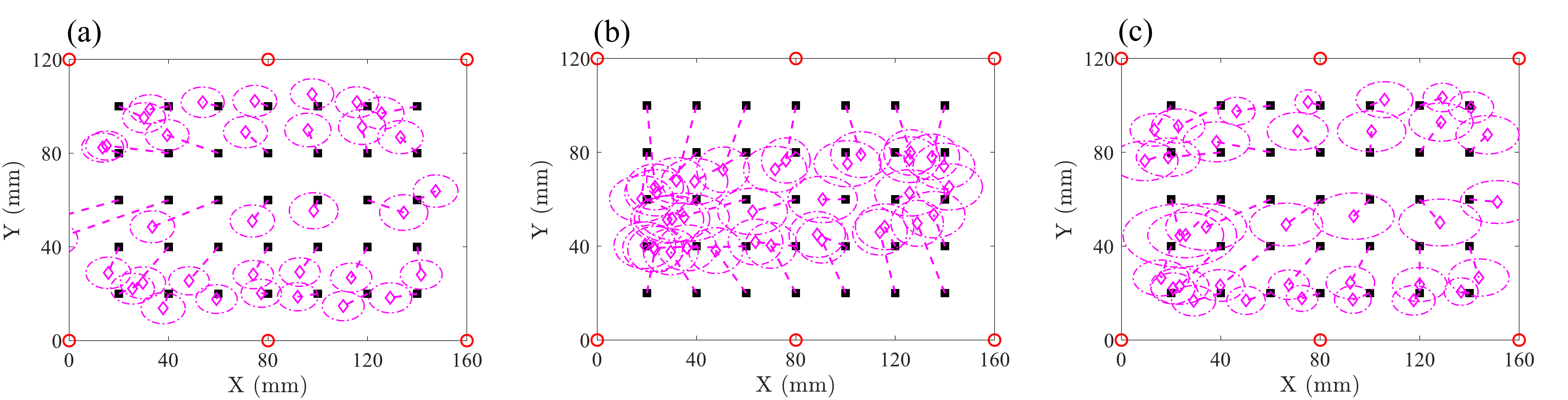}
	\caption{Interpolated localisation of TEM impacts with 9 RI using three different kernels: (a) COS kernel, (b) RBF kernel, and (c) COMP kernel. The COS kernel performs well in localising impacts in regions with larger TDOA values, while the RBF kernel excels in regions with smaller TDOA values. The COMP kernel effectively integrates the strengths of both COS and RBF kernels, achieving superior overall accuracy. }
	\label{FIG: fig_loc_9train_diff_kernels}
\end{figure}

\cref{FIG: fig_loc_9train_diff_kernels} illustrates the impact localisation of TEM impacts with 9 RI using three different kernels: COS, RBF, and COMP. The COS kernel performs poorly in regions with smaller TDOA values but excels in regions with larger TDOA values. Conversely, the RBF kernel performs well in regions with smaller TDOA values but struggles in areas with larger TDOA values. The COMP kernel effectively integrates the strengths of both COS and RBF kernels, achieving consistently high accuracy across all regions.

The above analysis underscores the robustness and effectiveness of the proposed kernel design and fusion strategy, offering a reliable and adaptable solution for accurate impact localisation under diverse uncertainties and varying numbers of reference impacts.

\subsection{Impact localisation: interpolation (int) vs. extrapolation (ext)}
\begin{figure}[htb] 
	\centering
		\includegraphics[width=1\columnwidth]{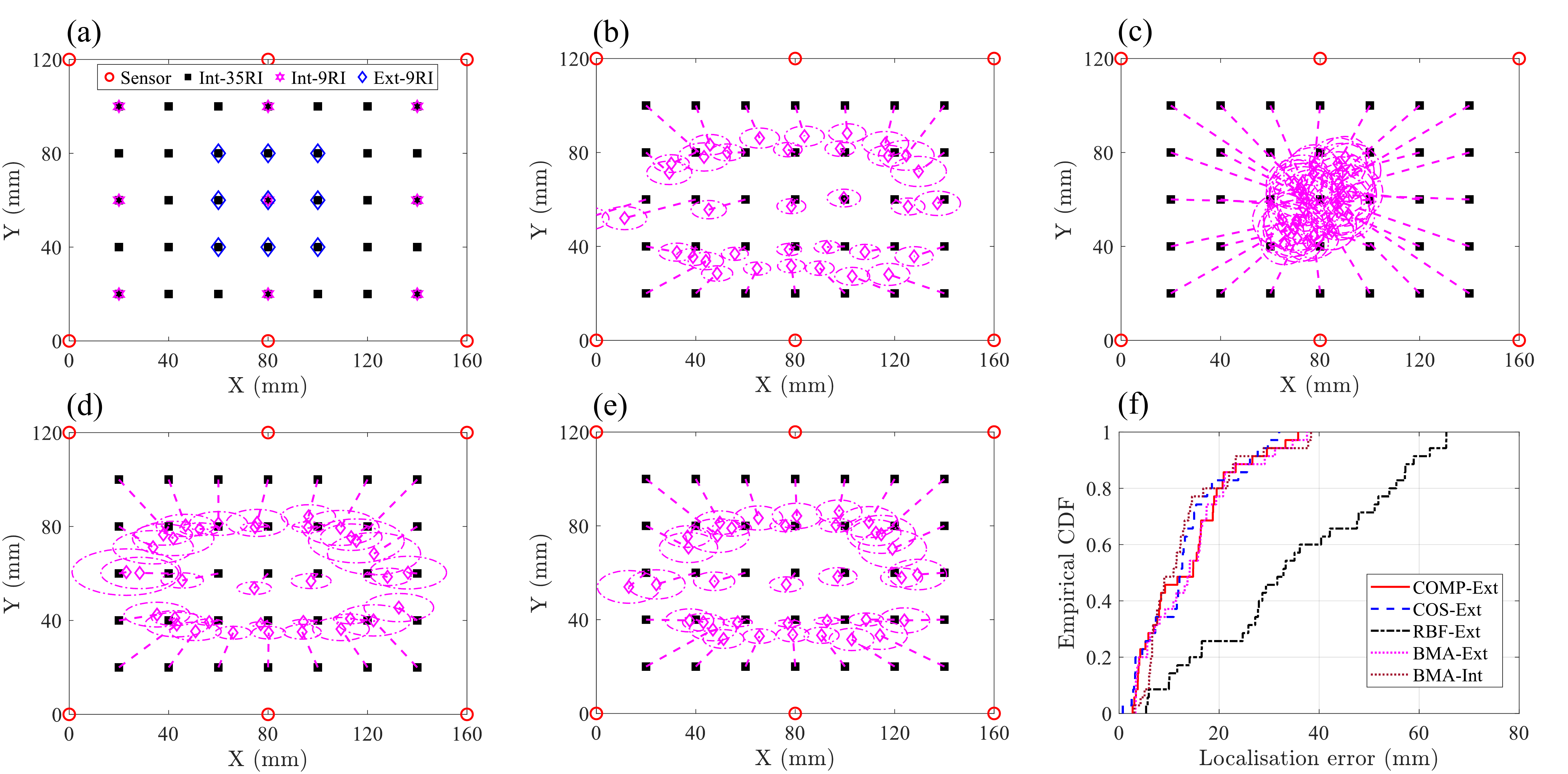}
	\caption{Extrapolated localisation of TEM impacts with the 9 innermost reference impacts (RI), referred to as 'Ext-9RI': (a) illustration of the RI set for extrapolation, denoted as 'Ext-9RI', (b) localisation results using the COS kernel, (c) localisation results using the RBF kernel, (d) localisation results using the COMP kernel, (e) localisation results using BMA fusion, (f) empirical CDF of localisation errors for the 'Ext-9RI' reference set.}
	\label{FIG: fig_loc_interp_vs_extrap}
\end{figure}

The above impact localisation represents an interpolation problem, where all target (testing) impacts lie within the coverage area of the reference impacts. To assess the performance of the proposed impact localisation method in handling extrapolation scenarios, another subset of 9 innermost reference impacts (RI) was constructed, referred to as 'Ext-9RI'. As shown in \cref{FIG: fig_loc_interp_vs_extrap}(a), compared to the 'Int-9RI' set (denoted by magenta hexagrams), which forms a rectangular region covering the entire 35 RIs, the 'Ext-9RI' set (denoted by blue diamonds) forms a smaller rectangle. Consequently, impacts located outside this smaller rectangle are considered part of an extrapolation problem.

\cref{FIG: fig_loc_interp_vs_extrap}(b)-(e) illustrates the extrapolated impact localisation using the 'Ext-9RI' reference set with three different kernels and BMA fusion. Combined with the empirical CDF of localisation errors shown in \cref{FIG: fig_loc_interp_vs_extrap}(f), it is evident that the RBF kernel struggles with extrapolation, failing to localise impacts outside the coverage area of the training reference impacts and resulting in substantial localisation errors and variances.

In contrast, the COS and COMP kernels exhibit some extrapolation capability, primarily due to their use of the TDOA order, which allows them to make reasonable predictions for impacts outside the reference coverage. Furthermore, the BMA fusion effectively combines the predictions of all three kernels, assigning greater weights to the COS and COMP kernels due to their lower predictive variances. As a result, BMA inherits the extrapolation ability of the COS and COMP kernels, offering improved localisation performance in extrapolation scenarios.

Compared to the interpolation set 'Int-9RI', which also consists of 9 reference impacts, the extrapolation set 'Ext-9RI' exhibits reduced overall localisation accuracy. The 'Ext-9RI' set achieves a mean localisation error of 14.2 mm, compared to 12.9 mm for the 'Int-9RI' set. This result confirms the advantage of surrogate modelling in handling interpolation problems over extrapolation scenarios. For data-driven impact localisation, it is recommended that the reference impacts encompass the potential locations of external impact events, ensuring that localisation tasks remain interpolation problems, which yield higher accuracy.

\section{Conclusion and outlook}  \label{section Conclusion and outlook}
This study introduced a robust data-driven approach for impact localisation on composite aircraft structures through kernel design and Bayesian fusion, effectively addressing environmental and operational uncertainties. With four passive sensors and nine reference impacts, the proposed method achieves a mean localisation error of approximately 12.9 mm under temperature variations of 46 $^\circ C$, corresponding to less than 4.5$\%$ of the characteristic structural dimension. This demonstrates its high accuracy and reliability without the need for explicit temperature compensation. Furthermore, increasing the number of sensors and reference impacts further improves localisation accuracy, reinforcing the adaptability of the approach to varying monitoring configurations. The key conclusions are summarised as follows:

\begin{itemize}
    \item Invariant TDOA order: The order of TDOAs between sensors remains consistent under variations in temperature and TDOA extraction frequency, forming the basis for the development of a composite kernel for GPR by incorporating TDOA order invariance.

    \item Kernel design: The integration of distance-based (RBF) and order-based (COS) kernels enables robust handling of environmental and operational variability, significantly improving localisation accuracy.

    \item Bayesian kernel fusion: The use of Bayesian model averaging to integrate multiple kernels ensures adaptive performance by balancing prediction accuracy and uncertainty quantification.
    
    \item Experimental validation: Extensive testing under varying conditions, including impact mass, energy, angle, and temperature—demonstrates the method’s generalisability and practical applicability, outperforming existing uni-kernel techniques.

    \item Pre-processing techniques: Sample standardisation of TDOA inputs proves to be a vital preprocessing step, preserving internal consistency and enhancing compatibility with GPR models.

    \item Independence from boundary conditions: As the impact wavefront is independent of boundary conditions, the proposed TDOA-based impact localisation method demonstrates potential effectiveness irrespective of structural boundaries.

    \item Interpolation over extrapolation: While the proposed composite kernel demonstrates the ability to localise impacts in extrapolation scenarios, experimental results confirm the superiority of surrogate modelling in handling interpolation over extrapolation. For data-driven impact localisation using surrogate models, it is advisable to ensure that reference impacts are strategically placed to cover potential external impact locations, thereby maximising localisation accuracy and reliability.
\end{itemize}

Beyond its strong performance on flat composite panels, the proposed method holds significant potential for scaling up to more complex aerostructures, including curved and three-dimensional components, where anisotropic wave propagation and structural discontinuities pose additional challenges. The probabilistic nature of the GPR framework enables the seamless incorporation of additional uncertainties, such as variations in structural stiffness, sensor noise, and complex boundary conditions, making it highly versatile for real-world applications.

Future research will focus on optimising the approach for more complex aerostructures with sparse sensor networks, reducing computational costs for real-time localisation, and integrating advanced deep learning techniques to further enhance accuracy and efficiency. Additionally, the development of hybrid physics-informed machine learning models will be explored, combining data-driven insights with fundamental wave propagation principles. This integration will enhance the robustness, scalability, and extrapolation capability of the method, ensuring its effectiveness for SHM applications in complex and uncertain operational environments.

\section*{CRediT authorship contribution statement}
\textbf{Dong Xiao}: Conceptualization, Methodology, Software, Data curation, Formal analysis, Writing - Original draft preparation, Writing - Review Editing. \textbf{Zahra Sharif-Khodaei}: Conceptualization, Methodology, Writing - Review Editing, Supervision. \textbf{M. H. Aliabadi}: Conceptualization, Methodology, Writing - Review Editing, Supervision.

\section*{Declaration of competing interest}
The authors declare that they have no known competing financial interests or personal relationships that could have appeared to influence the work reported in this paper.

\section*{Acknowledgements}

The first author gratefully acknowledges the China Scholarship Council (CSC) for supporting his Ph.D. studies through a scholarship (No. [2021]339). The authors also extend their gratitude to Dr. Aldyandra Hami Seno\href{https://orcid.org/0000-0001-9945-5299}{\includegraphics[scale=0.04]{orcidicon.pdf}}, a former Ph.D. member of the research group, for generating the guided drop mass experimental data under temperature variations (TEM), which were utilised in this study.

\section*{Data availability}
Data will be made available on request.



\begin{thebibliography}{70}
\providecommand{\natexlab}[1]{#1}
\providecommand{\url}[1]{\texttt{#1}}
\expandafter\ifx\csname urlstyle\endcsname\relax
  \providecommand{\doi}[1]{doi: #1}\else
  \providecommand{\doi}{doi: \begingroup \urlstyle{rm}\Url}\fi

\bibitem[Vincent and Emilie(2011)]{vincent_damage_2011}
Faivre Vincent and Morteau Emilie.
\newblock Damage tolerant composite fuselage sizing, characterization of accidental damage threat.
\newblock \emph{Airbus Tech Mag (FAST)}, 48:\penalty0 10--16, 2011.

\bibitem[Davies and Olsson(2004)]{davies_impact_2004}
G.~a.~O. Davies and R.~Olsson.
\newblock Impact on composite structures.
\newblock \emph{The Aeronautical Journal}, 108\penalty0 (1089):\penalty0 541--563, November 2004.
\newblock ISSN 0001-9240, 2059-6464.
\newblock \doi{10.1017/S0001924000000385}.

\bibitem[Dvorak and Laws(1987)]{dvorak_analysis_1987}
George~J. Dvorak and Norman Laws.
\newblock Analysis of {Progressive} {Matrix} {Cracking} {In} {Composite} {Laminates} {II}. {First} {Ply} {Failure}.
\newblock \emph{Journal of Composite Materials}, 21\penalty0 (4):\penalty0 309--329, April 1987.
\newblock ISSN 0021-9983.
\newblock \doi{10.1177/002199838702100402}.
\newblock Publisher: SAGE Publications Ltd STM.

\bibitem[Olsson et~al.(2006)Olsson, Donadon, and Falzon]{olsson_delamination_2006}
Robin Olsson, Mauricio~V. Donadon, and Brian~G. Falzon.
\newblock Delamination threshold load for dynamic impact on plates.
\newblock \emph{International Journal of Solids and Structures}, 43\penalty0 (10):\penalty0 3124--3141, May 2006.
\newblock ISSN 0020-7683.
\newblock \doi{10.1016/j.ijsolstr.2005.05.005}.

\bibitem[Shah et~al.(2019)Shah, Karuppanan, Megat-Yusoff, and Sajid]{shah_impact_2019}
S.~Z.~H. Shah, S.~Karuppanan, P.~S.~M. Megat-Yusoff, and Z.~Sajid.
\newblock Impact resistance and damage tolerance of fiber reinforced composites: {A} review.
\newblock \emph{Composite Structures}, 217:\penalty0 100--121, June 2019.
\newblock ISSN 0263-8223.
\newblock \doi{10.1016/j.compstruct.2019.03.021}.

\bibitem[Seydel and Chang(2001)]{seydel_impact_2001}
Robert Seydel and Fu-Kuo Chang.
\newblock Impact identification of stiffened composite panels: {I}. {System} development.
\newblock \emph{Smart Materials and Structures}, 10\penalty0 (2):\penalty0 354--369, 2001.
\newblock ISSN 0964-1726.
\newblock \doi{10.1088/0964-1726/10/2/323}.

\bibitem[Yu et~al.(2023)Yu, Xu, Sun, and Du]{yu_impact_2023}
Zexing Yu, Chao Xu, Jiaying Sun, and Fei Du.
\newblock Impact localization and force reconstruction for composite plates based on virtual time reversal processing with time-domain spectral finite element method.
\newblock \emph{Structural Health Monitoring}, 22\penalty0 (6):\penalty0 4149--4170, November 2023.
\newblock ISSN 1475-9217.
\newblock \doi{10.1177/14759217231164297}.

\bibitem[Liu and Wang(2023)]{liu_quantification_2023}
Yaru Liu and Lei Wang.
\newblock Quantification, localization, and reconstruction of impact force on interval composite structures.
\newblock \emph{International Journal of Mechanical Sciences}, 239:\penalty0 107873, February 2023.
\newblock ISSN 0020-7403.
\newblock \doi{10.1016/j.ijmecsci.2022.107873}.

\bibitem[Xiao et~al.(2024{\natexlab{a}})Xiao, Khodaei, and Aliabadi]{xiao_impact_2024}
Dong Xiao, Zahra~Sharif Khodaei, and M~H~Ferri Aliabadi.
\newblock Impact {Identification} {Based} on {Surrogate}-assisted {Efficient} {Global} {Optimisation}.
\newblock \emph{Procedia Structural Integrity}, 52:\penalty0 667--678, January 2024{\natexlab{a}}.
\newblock ISSN 2452-3216.
\newblock \doi{10.1016/j.prostr.2023.12.067}.

\bibitem[Yu et~al.(2024)Yu, Sun, Yue, and Xu]{yu_accelerated_2024}
Zexing Yu, Jiaying Sun, Maoling Yue, and Chao Xu.
\newblock Accelerated strategies for impact identification of composite laminate using sparse sensor networks and model-based virtual time reversal.
\newblock \emph{Structural Health Monitoring}, page 14759217241290260, November 2024.
\newblock ISSN 1475-9217.
\newblock \doi{10.1177/14759217241290260}.

\bibitem[Goutaudier et~al.(2020{\natexlab{a}})Goutaudier, Gendre, Kehr-Candille, and Ohayon]{goutaudier_impulse_2020}
Dimitri Goutaudier, Didier Gendre, Véronique Kehr-Candille, and Roger Ohayon.
\newblock Impulse identification technique by estimating specific modal ponderations from vibration measurements.
\newblock \emph{Journal of Sound and Vibration}, 474:\penalty0 115263, May 2020{\natexlab{a}}.
\newblock ISSN 0022-460X.
\newblock \doi{10.1016/j.jsv.2020.115263}.

\bibitem[Goutaudier et~al.(2020{\natexlab{b}})Goutaudier, Gendre, Kehr-Candille, and Ohayon]{goutaudier_single-sensor_2020}
Dimitri Goutaudier, Didier Gendre, Véronique Kehr-Candille, and Roger Ohayon.
\newblock Single-sensor approach for impact localization and force reconstruction by using discriminating vibration modes.
\newblock \emph{Mechanical Systems and Signal Processing}, 138:\penalty0 106534, April 2020{\natexlab{b}}.
\newblock ISSN 0888-3270.
\newblock \doi{10.1016/j.ymssp.2019.106534}.

\bibitem[Zhang et~al.(2024)Zhang, Liu, Hong, Wang, Zhou, and Liao]{zhang_efficient_2024}
Li~Zhang, Mingyao Liu, Liu Hong, Zechao Wang, Zude Zhou, and Wenlin Liao.
\newblock An efficient impact force identification methodology via a single sensor utilizing the concept of generalized transmissibility.
\newblock \emph{Mechanical Systems and Signal Processing}, 211:\penalty0 111222, April 2024.
\newblock ISSN 0888-3270.
\newblock \doi{10.1016/j.ymssp.2024.111222}.

\bibitem[El-Bakari et~al.(2014)El-Bakari, Khamlichi, Jacquelin, and Dkiouak]{el-bakari_assessing_2014}
Abdelali El-Bakari, Abdellatif Khamlichi, Eric Jacquelin, and Rachid Dkiouak.
\newblock Assessing impact force localization by using a particle swarm optimization algorithm.
\newblock \emph{Journal of Sound and Vibration}, 333\penalty0 (6):\penalty0 1554--1561, March 2014.
\newblock ISSN 0022-460X.
\newblock \doi{10.1016/j.jsv.2013.11.032}.

\bibitem[Atobe et~al.(2014)Atobe, Sugimoto, Hu, and Fukunaga]{atobe_impact_2014}
Satoshi Atobe, Sunao Sugimoto, Ning Hu, and Hisao Fukunaga.
\newblock Impact damage monitoring of {FRP} pressure vessels based on impact force identification.
\newblock \emph{Advanced Composite Materials}, 23\penalty0 (5-6):\penalty0 491--505, September 2014.
\newblock ISSN 0924-3046.
\newblock \doi{10.1080/09243046.2014.915112}.
\newblock Publisher: Taylor \& Francis.

\bibitem[Ciampa and Meo(2012)]{ciampa_impact_2012}
F~Ciampa and M~Meo.
\newblock Impact detection in anisotropic materials using a time reversal approach.
\newblock \emph{Structural Health Monitoring}, 11\penalty0 (1):\penalty0 43--49, 2012.
\newblock ISSN 1475-9217.
\newblock \doi{10.1177/1475921710395815}.
\newblock Publisher: SAGE Publications.

\bibitem[Simone et~al.(2019)Simone, Ciampa, and Meo]{simone_hierarchical_2019}
Mario Emanuele~De Simone, Francesco Ciampa, and Michele Meo.
\newblock A hierarchical method for the impact force reconstruction in composite structures.
\newblock \emph{Smart Materials and Structures}, 28\penalty0 (8):\penalty0 085022, 2019.
\newblock ISSN 0964-1726.
\newblock \doi{10.1088/1361-665X/aae11c}.

\bibitem[Chen et~al.(2012)Chen, Li, and Yuan]{chen_impact_2012}
Chunlin Chen, Yulong Li, and Fuh-Gwo Yuan.
\newblock Impact source identification in finite isotropic plates using a time-reversal method: experimental study.
\newblock \emph{Smart Materials and Structures}, 21\penalty0 (10):\penalty0 105025, August 2012.
\newblock ISSN 0964-1726.
\newblock \doi{10.1088/0964-1726/21/10/105025}.

\bibitem[Liu and Li(2018)]{liu_novel_2018}
Jie Liu and Bing Li.
\newblock A novel strategy for response and force reconstruction under impact excitation.
\newblock \emph{Journal of Mechanical Science and Technology}, 32\penalty0 (8):\penalty0 3581--3596, August 2018.
\newblock ISSN 1976-3824.
\newblock \doi{10.1007/s12206-018-0709-4}.

\bibitem[Giammaria et~al.(2023)Giammaria, Del~Bianco, Raponi, Fiumarella, Ciardiello, Boria, Duddeck, and Belingardi]{giammaria_material_2023}
Valentina Giammaria, Giulia Del~Bianco, Elena Raponi, Dario Fiumarella, Raffaele Ciardiello, Simonetta Boria, Fabian Duddeck, and Giovanni Belingardi.
\newblock Material parameter optimization of flax/epoxy composite laminates under low-velocity impact.
\newblock \emph{Composite Structures}, 321:\penalty0 117303, October 2023.
\newblock ISSN 0263-8223.
\newblock \doi{10.1016/j.compstruct.2023.117303}.

\bibitem[Xu et~al.(2023)Xu, Wang, Hu, Du, Du, Yu, and Lei]{xu_fem-based_2023}
Tian Xu, Zhen Wang, Yingda Hu, Shilun Du, Ao~Du, Zhenyang Yu, and Yong Lei.
\newblock A {FEM}-based direct method for identification of {Young}’s modulus and boundary conditions in three-dimensional linear elasticity from local observation.
\newblock \emph{International Journal of Mechanical Sciences}, 237:\penalty0 107797, January 2023.
\newblock ISSN 0020-7403.
\newblock \doi{10.1016/j.ijmecsci.2022.107797}.

\bibitem[Ereiz et~al.(2022)Ereiz, Duvnjak, and Fernando Jiménez-Alonso]{ereiz_review_2022}
Suzana Ereiz, Ivan Duvnjak, and Javier Fernando Jiménez-Alonso.
\newblock Review of finite element model updating methods for structural applications.
\newblock \emph{Structures}, 41:\penalty0 684--723, July 2022.
\newblock ISSN 2352-0124.
\newblock \doi{10.1016/j.istruc.2022.05.041}.

\bibitem[Szabó and Babuška(2021)]{szabo_finite_2021}
Barna Szabó and Ivo Babuška.
\newblock \emph{Finite {Element} {Analysis}: {Method}, {Verification} and {Validation}}.
\newblock John Wiley \& Sons, June 2021.
\newblock ISBN 978-1-119-42642-4.

\bibitem[Sharif-Khodaei et~al.(2012)Sharif-Khodaei, Ghajari, and Aliabadi]{sharif-khodaei_determination_2012}
Z.~Sharif-Khodaei, M.~Ghajari, and M.~H. Aliabadi.
\newblock Determination of impact location on composite stiffened panels.
\newblock \emph{Smart Materials and Structures}, 21\penalty0 (10):\penalty0 105026, 2012.
\newblock ISSN 0964-1726.
\newblock \doi{10.1088/0964-1726/21/10/105026}.

\bibitem[Ghajari et~al.(2013)Ghajari, Sharif-Khodaei, Aliabadi, and Apicella]{ghajari_identification_2013}
M.~Ghajari, Z.~Sharif-Khodaei, M.~H. Aliabadi, and A.~Apicella.
\newblock Identification of impact force for smart composite stiffened panels.
\newblock \emph{Smart Materials and Structures}, 22\penalty0 (8):\penalty0 085014, 2013.
\newblock ISSN 0964-1726.
\newblock \doi{10.1088/0964-1726/22/8/085014}.

\bibitem[Seno and Aliabadi(2021)]{seno_uncertainty_2021}
Aldyandra~Hami Seno and MH~Ferri Aliabadi.
\newblock Uncertainty quantification for impact location and force estimation in composite structures.
\newblock \emph{Structural Health Monitoring}, page 147592172110202, June 2021.
\newblock ISSN 1475-9217, 1741-3168.
\newblock \doi{10.1177/14759217211020255}.

\bibitem[Jones et~al.(2022)Jones, Rogers, Worden, and Cross]{jones_bayesian_2022}
M.~R. Jones, T.~J. Rogers, K.~Worden, and E.~J. Cross.
\newblock A {Bayesian} methodology for localising acoustic emission sources in complex structures.
\newblock \emph{Mechanical Systems and Signal Processing}, 163:\penalty0 108143, January 2022.
\newblock ISSN 0888-3270.
\newblock \doi{10.1016/j.ymssp.2021.108143}.

\bibitem[Grasboeck et~al.(2024)Grasboeck, Humer, and Benjeddou]{grasboeck_detection_2024}
Lukas Grasboeck, Alexander Humer, and Ayech Benjeddou.
\newblock Detection and time-of-arrival estimation of impact-induced waves in composite laminates.
\newblock \emph{Structural Health Monitoring}, page 14759217241260848, July 2024.
\newblock ISSN 1475-9217.
\newblock \doi{10.1177/14759217241260848}.
\newblock Publisher: SAGE Publications.

\bibitem[Zeng et~al.(2025)Zeng, Deng, Yang, Yang, Yang, and Wu]{zeng_hybrid_2025}
Xu~Zeng, Deshuang Deng, Hongjuan Yang, Zhengyan Yang, Lei Yang, and Zhanjun Wu.
\newblock Hybrid {DoA}-{TDoA} method for impact localization on thin-walled structures using sensor clusters.
\newblock \emph{IEEE Sensors Journal}, pages 1--1, 2025.
\newblock ISSN 1558-1748.
\newblock \doi{10.1109/JSEN.2024.3515461}.
\newblock Conference Name: IEEE Sensors Journal.

\bibitem[Zhu et~al.(2018)Zhu, Qing, and Liu]{zhu_two-step_2018}
Kaige Zhu, Xinlin~P. Qing, and Bin Liu.
\newblock A two-step impact localization method for composite structures with a parameterized laminate model.
\newblock \emph{Composite Structures}, 192:\penalty0 500--506, May 2018.
\newblock ISSN 0263-8223.
\newblock \doi{10.1016/j.compstruct.2018.03.052}.

\bibitem[Zhu et~al.(2020)Zhu, Qing, Liu, and He]{zhu_passive_2020}
Kaige Zhu, Xinlin~P. Qing, Bin Liu, and Tiren He.
\newblock A passive localization method for stiffened composite structures with a parameterized laminate model.
\newblock \emph{Journal of Sound and Vibration}, 489:\penalty0 115683, December 2020.
\newblock ISSN 0022-460X.
\newblock \doi{10.1016/j.jsv.2020.115683}.

\bibitem[Meo et~al.(2005)Meo, Zumpano, Piggott, and Marengo]{meo_impact_2005}
M.~Meo, G.~Zumpano, M.~Piggott, and G.~Marengo.
\newblock Impact identification on a sandwich plate from wave propagation responses.
\newblock \emph{Composite Structures}, 71\penalty0 (3):\penalty0 302--306, December 2005.
\newblock ISSN 0263-8223.
\newblock \doi{10.1016/j.compstruct.2005.09.028}.

\bibitem[Xiao et~al.(2025)Xiao, Rodrigues, Sharif-Khodaei, and Aliabadi]{xiao_general_2025}
Dong Xiao, Francisco de~Sá Rodrigues, Zahra Sharif-Khodaei, and M.~H. Aliabadi.
\newblock A general probabilistic framework for impact localisation based on flexural wave propagation.
\newblock \emph{Mechanical Systems and Signal Processing}, 226:\penalty0 112320, March 2025.
\newblock ISSN 0888-3270.
\newblock \doi{10.1016/j.ymssp.2025.112320}.

\bibitem[Dehghan~Niri et~al.(2014)Dehghan~Niri, Farhidzadeh, and Salamone]{dehghan_niri_nonlinear_2014}
E.~Dehghan~Niri, A.~Farhidzadeh, and S.~Salamone.
\newblock Nonlinear {Kalman} {Filtering} for acoustic emission source localization in anisotropic panels.
\newblock \emph{Ultrasonics}, 54\penalty0 (2):\penalty0 486--501, February 2014.
\newblock ISSN 0041-624X.
\newblock \doi{10.1016/j.ultras.2013.07.016}.

\bibitem[Niri and Salamone(2012)]{niri_probabilistic_2012}
E.~Dehghan Niri and S.~Salamone.
\newblock A probabilistic framework for acoustic emission source localization in plate-like structures.
\newblock \emph{Smart Materials and Structures}, 21\penalty0 (3):\penalty0 035009, 2012.
\newblock ISSN 0964-1726.
\newblock \doi{10.1088/0964-1726/21/3/035009}.

\bibitem[Deng et~al.(2024)Deng, Zeng, Yang, Yang, Zhang, Ma, Xu, Yang, and Wu]{deng_multi-frequency_2024}
Deshuang Deng, Xu~Zeng, Zhengyan Yang, Yu~Yang, Sheng Zhang, Shuyi Ma, Hao Xu, Lei Yang, and Zhanjun Wu.
\newblock Multi-frequency probabilistic imaging fusion for impact localization on aircraft composite structures.
\newblock \emph{Structural Health Monitoring}, page 14759217241233181, March 2024.
\newblock ISSN 1475-9217.
\newblock \doi{10.1177/14759217241233181}.

\bibitem[Tabian et~al.(2019)Tabian, Fu, and Sharif~Khodaei]{tabian_convolutional_2019}
Iuliana Tabian, Hailing Fu, and Zahra Sharif~Khodaei.
\newblock A {Convolutional} {Neural} {Network} for {Impact} {Detection} and {Characterization} of {Complex} {Composite} {Structures}.
\newblock \emph{Sensors}, 19\penalty0 (22):\penalty0 4933, January 2019.
\newblock ISSN 1424-8220.
\newblock \doi{10.3390/s19224933}.

\bibitem[Zhao et~al.(2024)Zhao, Zhang, Liu, and Qing]{zhao_impact_2024}
Bowen Zhao, Yiliang Zhang, Qijian Liu, and Xinlin Qing.
\newblock Impact monitoring of large size complex metal structures based on sparse sensor array and transfer learning.
\newblock \emph{Ultrasonics}, 140:\penalty0 107305, May 2024.
\newblock ISSN 0041-624X.
\newblock \doi{10.1016/j.ultras.2024.107305}.

\bibitem[Zhou et~al.(2024{\natexlab{a}})Zhou, Qiao, Liu, Cheng, and Chen]{zhou_impact_2024}
Rui Zhou, Baijie Qiao, Junjiang Liu, Wei Cheng, and Xuefeng Chen.
\newblock Impact force localization and reconstruction via gated temporal convolutional network.
\newblock \emph{Aerospace Science and Technology}, 144:\penalty0 108819, January 2024{\natexlab{a}}.
\newblock ISSN 1270-9638.
\newblock \doi{10.1016/j.ast.2023.108819}.

\bibitem[Huang et~al.(2023{\natexlab{a}})Huang, Liao, Sun, Wang, and Qing]{huang_hybrid_2023}
Chenhui Huang, Weilin Liao, Hu~Sun, Yishou Wang, and Xinlin Qing.
\newblock A hybrid {FCN}-{BiGRU} with transfer learning for low-velocity impact identification on aircraft structure.
\newblock \emph{Smart Materials and Structures}, 32\penalty0 (5):\penalty0 055012, April 2023{\natexlab{a}}.
\newblock ISSN 0964-1726.
\newblock \doi{10.1088/1361-665X/acc623}.
\newblock Publisher: IOP Publishing.

\bibitem[Zhou et~al.(2024{\natexlab{b}})Zhou, Cai, Dong, Zhang, and Peng]{zhou_data-physics_2024}
Jiaming Zhou, Yinshan Cai, Longlei Dong, Bo~Zhang, and Zhike Peng.
\newblock Data-physics hybrid-driven deep learning method for impact force identification.
\newblock \emph{Mechanical Systems and Signal Processing}, 211:\penalty0 111238, April 2024{\natexlab{b}}.
\newblock ISSN 0888-3270.
\newblock \doi{10.1016/j.ymssp.2024.111238}.

\bibitem[Zhao and Chen(2023)]{zhao_spatial-temporal_2023}
Zhimin Zhao and Nian-Zhong Chen.
\newblock Spatial-temporal graph convolutional networks ({STGCN}) based method for localizing acoustic emission sources in composite panels.
\newblock \emph{Composite Structures}, 323:\penalty0 117496, November 2023.
\newblock ISSN 0263-8223.
\newblock \doi{10.1016/j.compstruct.2023.117496}.

\bibitem[Huang et~al.(2023{\natexlab{b}})Huang, Tao, Ji, and Qiu]{huang_impact_2023}
Chun Huang, Chongcong Tao, Hongli Ji, and Jinhao Qiu.
\newblock Impact force reconstruction and localization using {Distance}-assisted {Graph} {Neural} {Network}.
\newblock \emph{Mechanical Systems and Signal Processing}, 200:\penalty0 110606, October 2023{\natexlab{b}}.
\newblock ISSN 0888-3270.
\newblock \doi{10.1016/j.ymssp.2023.110606}.

\bibitem[Seno et~al.(2019)Seno, Sharif~Khodaei, and Aliabadi]{seno_passive_2019}
Aldyandra~Hami Seno, Zahra Sharif~Khodaei, and M.~H.~Ferri Aliabadi.
\newblock Passive sensing method for impact localisation in composite plates under simulated environmental and operational conditions.
\newblock \emph{Mechanical Systems and Signal Processing}, 129:\penalty0 20--36, 2019.
\newblock ISSN 0888-3270.
\newblock \doi{10.1016/j.ymssp.2019.04.023}.

\bibitem[Seno and Aliabadi(2019)]{seno_impact_2019}
Aldyandra~Hami Seno and M.~H.~Ferri Aliabadi.
\newblock Impact {Localisation} in {Composite} {Plates} of {Different} {Stiffness} {Impactors} under {Simulated} {Environmental} and {Operational} {Conditions}.
\newblock \emph{Sensors}, 19\penalty0 (17):\penalty0 3659, January 2019.
\newblock ISSN 1424-8220.
\newblock \doi{10.3390/s19173659}.

\bibitem[Hami~Seno and Ferri~Aliabadi(2023)]{hami_seno_multifidelity_2023}
Aldyandra Hami~Seno and M.~H. Ferri~Aliabadi.
\newblock Multifidelity data augmentation for data driven passive impact location and force estimation in composite structures under simulated environmental and operational conditions.
\newblock \emph{Mechanical Systems and Signal Processing}, 195:\penalty0 110288, July 2023.
\newblock ISSN 0888-3270.
\newblock \doi{10.1016/j.ymssp.2023.110288}.

\bibitem[Kalimullah et~al.(2023)Kalimullah, Shelke, and Habib]{kalimullah_probabilistic_2023}
Nur M.~M. Kalimullah, Amit Shelke, and Anowarul Habib.
\newblock A probabilistic framework for source localization in anisotropic composite using transfer learning based multi-fidelity physics informed neural network ({mfPINN}).
\newblock \emph{Mechanical Systems and Signal Processing}, 197:\penalty0 110360, August 2023.
\newblock ISSN 0888-3270.
\newblock \doi{10.1016/j.ymssp.2023.110360}.

\bibitem[Xiao et~al.(2024{\natexlab{b}})Xiao, Sharif-Khodaei, and Aliabadi]{xiao_hybrid_2024}
Dong Xiao, Zahra Sharif-Khodaei, and M.~H. Aliabadi.
\newblock Hybrid physics-based and data-driven impact localisation for composite laminates.
\newblock \emph{International Journal of Mechanical Sciences}, 274:\penalty0 109222, July 2024{\natexlab{b}}.
\newblock ISSN 0020-7403.
\newblock \doi{10.1016/j.ijmecsci.2024.109222}.

\bibitem[Aliabadi and Sharif~Khodaei(2018)]{aliabadi_structural_2018}
M~H~Ferri Aliabadi and Z~Sharif~Khodaei.
\newblock \emph{Structural {Health} {Monitoring} for {Advanced} {Composite} {Structures}}, volume~08 of \emph{Computational and {Experimental} {Methods} in {Structures}}.
\newblock World Scientific (Europe), February 2018.
\newblock ISBN 978-1-78634-392-5 978-1-78634-393-2.
\newblock \doi{10.1142/q0114}.

\bibitem[Moon(1973{\natexlab{a}})]{moon_critical_1973}
F.~C. Moon.
\newblock A critical survey of wave propagation and impact in composite materials.
\newblock Technical Report CR-121226, NASA, May 1973{\natexlab{a}}.

\bibitem[Kim and Moon(1979)]{kim_impact_1979}
Byoung~Sung Kim and Francis Moon.
\newblock Impact {Induced} {Stress} {Waves} in an {Anisotropic} {Plate}.
\newblock \emph{AIAA Journal}, 17\penalty0 (10):\penalty0 1126--1133, October 1979.
\newblock ISSN 0001-1452, 1533-385X.
\newblock \doi{10.2514/3.61287}.

\bibitem[Fu et~al.(2019)Fu, Sharif~Khodaei, and Aliabadi]{fu_event-triggered_2019}
Hailing Fu, Zahra Sharif~Khodaei, and M.~H.~Ferri Aliabadi.
\newblock An {Event}-{Triggered} {Energy}-{Efficient} {Wireless} {Structural} {Health} {Monitoring} {System} for {Impact} {Detection} in {Composite} {Airframes}.
\newblock \emph{IEEE Internet of Things Journal}, 6\penalty0 (1):\penalty0 1183--1192, 2019.
\newblock ISSN 2327-4662.
\newblock \doi{10.1109/JIOT.2018.2867722}.
\newblock Conference Name: IEEE Internet of Things Journal.

\bibitem[Abrate(1998)]{abrate_impact_1998}
Serge Abrate.
\newblock \emph{Impact on {Composite} {Structures}}.
\newblock Cambridge University Press, Cambridge, 1998.
\newblock ISBN 978-0-521-47389-7.
\newblock \doi{10.1017/CBO9780511574504}.

\bibitem[Moon(1973{\natexlab{b}})]{moon_theoretical_1973}
F.~C. Moon.
\newblock Theoretical analysis of impact in composite plates.
\newblock Technical Report AMS-1099, NASA, January 1973{\natexlab{b}}.

\bibitem[Daniel et~al.(1979)Daniel, Liber, and LaBedz]{daniel_wave_1979}
I.~M. Daniel, T.~Liber, and R.~H. LaBedz.
\newblock Wave propagation in transversely impacted composite laminates.
\newblock \emph{Experimental Mechanics}, 19\penalty0 (1):\penalty0 9, 1979.
\newblock ISSN 0014-4851.
\newblock \doi{10.1007/BF02327764}.

\bibitem[Tan and Sun(1982)]{tan_wave_1982}
T.~M. Tan and C.~T. Sun.
\newblock Wave propagation in graphite/epoxy laminates due to impact.
\newblock Technical Report NASA-CR-168057, NASA, December 1982.

\bibitem[Kundu(2014)]{kundu_acoustic_2014}
Tribikram Kundu.
\newblock Acoustic source localization.
\newblock \emph{Ultrasonics}, 54\penalty0 (1):\penalty0 25--38, January 2014.
\newblock ISSN 0041-624X.
\newblock \doi{10.1016/j.ultras.2013.06.009}.

\bibitem[Xiao et~al.(2024{\natexlab{c}})Xiao, Sharif-Khodaei, and Aliabadi]{xiao_impact_2024-1}
Dong Xiao, Zahra Sharif-Khodaei, and M.~H. Aliabadi.
\newblock Impact force identification for composite structures using adaptive wavelet-regularised deconvolution.
\newblock \emph{Mechanical Systems and Signal Processing}, 220:\penalty0 111608, November 2024{\natexlab{c}}.
\newblock ISSN 0888-3270.
\newblock \doi{10.1016/j.ymssp.2024.111608}.

\bibitem[Olsson(2000)]{olsson_mass_2000}
R.~Olsson.
\newblock Mass criterion for wave controlled impact response of composite plates.
\newblock \emph{Composites Part A: Applied Science and Manufacturing}, 31\penalty0 (8):\penalty0 879--887, August 2000.
\newblock ISSN 1359-835X.
\newblock \doi{10.1016/S1359-835X(00)00020-8}.

\bibitem[Schulz et~al.(2018)Schulz, Speekenbrink, and Krause]{schulz_tutorial_2018}
Eric Schulz, Maarten Speekenbrink, and Andreas Krause.
\newblock A tutorial on {Gaussian} process regression: {Modelling}, exploring, and exploiting functions.
\newblock \emph{Journal of Mathematical Psychology}, 85:\penalty0 1--16, August 2018.
\newblock ISSN 0022-2496.
\newblock \doi{10.1016/j.jmp.2018.03.001}.

\bibitem[Forrester et~al.(2008)Forrester, Sbester, and Keane]{forrester_engineering_2008}
Alexander I.~J. Forrester, Andrs Sbester, and Andy~J. Keane.
\newblock \emph{Engineering {Design} via {Surrogate} {Modelling}}.
\newblock John Wiley \& Sons, Ltd, Chichester, UK, July 2008.
\newblock ISBN 978-0-470-77080-1 978-0-470-06068-1.
\newblock \doi{10.1002/9780470770801}.

\bibitem[Satria~Palar et~al.(2020)Satria~Palar, Rizki~Zuhal, and Shimoyama]{satria_palar_gaussian_2020}
Pramudita Satria~Palar, Lavi Rizki~Zuhal, and Koji Shimoyama.
\newblock Gaussian {Process} {Surrogate} {Model} with {Composite} {Kernel} {Learning} for {Engineering} {Design}.
\newblock \emph{AIAA Journal}, 58\penalty0 (4):\penalty0 1864--1880, 2020.
\newblock ISSN 0001-1452.
\newblock \doi{10.2514/1.J058807}.
\newblock Publisher: American Institute of Aeronautics and Astronautics \_eprint: https://doi.org/10.2514/1.J058807.

\bibitem[Nabati et~al.(2022)Nabati, Ghorashi, and Shahbazian]{nabati_jgpr_2022}
Mohammad Nabati, Seyed~Ali Ghorashi, and Reza Shahbazian.
\newblock {JGPR}: a computationally efficient multi-target {Gaussian} process regression algorithm.
\newblock \emph{Machine Learning}, 111\penalty0 (6):\penalty0 1987--2010, June 2022.
\newblock ISSN 1573-0565.
\newblock \doi{10.1007/s10994-022-06170-3}.

\bibitem[Gardner et~al.(2018)Gardner, Pleiss, Weinberger, Bindel, and Wilson]{gardner_gpytorch_2018}
Jacob Gardner, Geoff Pleiss, Kilian~Q Weinberger, David Bindel, and Andrew~G Wilson.
\newblock {GPyTorch}: {Blackbox} {Matrix}-{Matrix} {Gaussian} {Process} {Inference} with {GPU} {Acceleration}.
\newblock In \emph{Advances in {Neural} {Information} {Processing} {Systems}}, volume~31, pages 1--11. Curran Associates, Inc., 2018.

\bibitem[Paszke et~al.(2019)Paszke, Gross, Massa, Lerer, Bradbury, Chanan, Killeen, Lin, Gimelshein, Antiga, Desmaison, Kopf, Yang, DeVito, Raison, Tejani, Chilamkurthy, Steiner, Fang, Bai, and Chintala]{paszke_pytorch_2019}
Adam Paszke, Sam Gross, Francisco Massa, Adam Lerer, James Bradbury, Gregory Chanan, Trevor Killeen, Zeming Lin, Natalia Gimelshein, Luca Antiga, Alban Desmaison, Andreas Kopf, Edward Yang, Zachary DeVito, Martin Raison, Alykhan Tejani, Sasank Chilamkurthy, Benoit Steiner, Lu~Fang, Junjie Bai, and Soumith Chintala.
\newblock {PyTorch}: {An} {Imperative} {Style}, {High}-{Performance} {Deep} {Learning} {Library}.
\newblock In \emph{Advances in {Neural} {Information} {Processing} {Systems}}, volume~32, pages 1--12. Curran Associates, Inc., 2019.

\bibitem[Wasserman(2000)]{wasserman_bayesian_2000}
Larry Wasserman.
\newblock Bayesian {Model} {Selection} and {Model} {Averaging}.
\newblock \emph{Journal of Mathematical Psychology}, 44\penalty0 (1):\penalty0 92--107, March 2000.
\newblock ISSN 0022-2496.
\newblock \doi{10.1006/jmps.1999.1278}.

\bibitem[Suemasu and Majima(1996)]{suemasu_multiple_1996}
Hiroshi Suemasu and Osamu Majima.
\newblock Multiple {Delaminations} and their {Severity} in {Circular} {Axisymmetric} {Plates} {Subjected} to {Transverse} {Loading}.
\newblock \emph{Journal of Composite Materials}, 30\penalty0 (4):\penalty0 441--453, March 1996.
\newblock ISSN 0021-9983.
\newblock \doi{10.1177/002199839603000402}.

\bibitem[Olsson(2001)]{olsson_analytical_2001}
Robin Olsson.
\newblock Analytical prediction of large mass impact damage in composite laminates.
\newblock \emph{Composites Part A: Applied Science and Manufacturing}, 32\penalty0 (9):\penalty0 1207--1215, September 2001.
\newblock ISSN 1359-835X.
\newblock \doi{10.1016/S1359-835X(01)00073-2}.

\bibitem[Thiene et~al.(2014)Thiene, Ghajari, Galvanetto, and Aliabadi]{thiene_effects_2014}
M.~Thiene, M.~Ghajari, U.~Galvanetto, and M.~H. Aliabadi.
\newblock Effects of the transfer function evaluation on the impact force reconstruction with application to composite panels.
\newblock \emph{Composite Structures}, 114:\penalty0 1--9, August 2014.
\newblock ISSN 0263-8223.
\newblock \doi{10.1016/j.compstruct.2014.03.055}.

\bibitem[Xiao et~al.(2024{\natexlab{d}})Xiao, Sharif-Khodaei, and Aliabadi]{xiao_time_2024}
Dong Xiao, Zahra Sharif-Khodaei, and M.~H. Aliabadi.
\newblock Time of {Arrival} {Extraction} for {Impact} {Localisation} on {Composite} {Structures}.
\newblock \emph{e-Journal of Nondestructive Testing}, 29\penalty0 (07), July 2024{\natexlab{d}}.
\newblock \doi{10.58286/29601}.

\end{thebibliography}

\small{

}

\end{document}